\documentclass[12pt]{article}

\usepackage{amsmath,amsfonts,amssymb}
\usepackage{graphicx}
\usepackage{psfrag}
\usepackage{enumerate}
\usepackage{subfigure}

\makeatletter \@addtoreset{equation}{section}

\makeatletter\renewcommand\section{\@startsection {section}{1}{\z@}%
                                   {-3.5ex \@plus -1ex \@minus -.2ex}%nn
                                   {2.3ex \@plus.2ex}%
                                   {\normalfont\large\bfseries}}
\renewcommand\subsection{\@startsection{subsection}{2}{\z@}%
                                     {-3.25ex\@plus -1ex \@minus -.2ex}%
                                     {1.5ex \@plus .2ex}%
                                     {\normalfont\bfseries}}

\newcommand{\be}{\begin{equation}}
\newcommand{\ee}{\end{equation}}
\newcommand{\beq}{\begin{eqnarray}}
\newcommand{\eeq}{\end{eqnarray}}
\newcommand{\bea}{\begin{eqnarray}}
\newcommand{\eea}{\end{eqnarray}}

\def\[{\left [}
\def\]{\right ]}
\def\({\left (}
\def\){\right )}

\def\r2{\sqrt{2}}

\def\la{\langle}
\def\ra{\rangle}
\def\ea{A^0}  %Exclusive area
\def\ev{I^0}  %Exclusive volume
\def\nev{I}   %nonexclusive vol
\def\D{\Delta} %conformal dimension
\def\dvol{C_d} %volume of unit d-sphere

\def\li{{\rm Li_2}}
\def\dtwo{{\rm D_2}}
\def\nd{\#_d}

\newcommand{\bbibitem}[1]{\bibitem{#1}\marginpar{#1}}

\def\Label#1{\label{#1}%
  \smash{\hbox to0pt{\raise1ex\hbox{\tiny[#1]}\hss}}}
\def\noLabels{\let\Label=\label}
\def\nobbibitem{\let\bbibitem=\bibitem}

\begin{document}
%\noLabels % uncomment for final production
%\nobbibitem % uncomment for final production

\begin{titlepage}

\vfil\

\begin{center}

{\Large{\bf A Conformal Field Theory for Eternal Inflation}}

\vspace{3mm}

Ben Freivogel\footnote{e-mail: freivogel@berkeley.edu
} and Matthew Kleban\footnote{e-mail: mk161@nyu.edu}
\\

\vspace{8mm}
{ \textit{$^{\rm 1}$ 
Berkeley Center for Theoretical Physics, Department of Physics\\
University of California, Berkeley, CA 94720-7300, USA\\
{\rm and} \\
Lawrence Berkeley National Laboratory, Berkeley, CA 94720-8162, USA
}}

\bigskip\medskip
\centerline{\it $^{\rm 2}$Center for Cosmology and Particle Physics}
\smallskip\centerline{\it Department of Physics, New York University}
\smallskip\centerline{\it 4 Washington Place, New York, NY 10003.}

\vfil

\end{center}
\setcounter{footnote}{0}

\begin{abstract}
\noindent

We study a statistical model defined by a conformally invariant
distribution of overlapping spheres in arbitrary dimension $d$. The model arises
as the asymptotic distribution of cosmic bubbles in $d+1$ dimensional
de Sitter space, and also as the asymptotic distribution of bubble collisions
with the domain wall of a fiducial ``observation bubble'' in $d+2$
dimensional de Sitter space.  In this note we calculate the 2-,3-, and
4-point correlation functions of exponentials of the ``bubble
number operator" analytically in $d=2$.  We find that these
correlators
are free of infrared divergences, covariant under the global conformal
group, charge conserving, and transform with
positive conformal dimensions that are related in a novel way to the
charge.  Although by themselves these operators probably do not define
a full-fledged conformal field theory, one can use the partition
function on a sphere to compute an approximate central charge in the
2D case.  The theory in any dimension has a noninteracting limit when the
nucleation rate of the bubbles in the bulk is very large.  The theory
in two dimensions is related to some models of continuum percolation,
but it is conformal for all values of the tunneling rate.

\end{abstract}

\vspace{0.5in}

\end{titlepage}

\renewcommand{\baselinestretch}{1.05}  

%\tableofcontents

%\newpage

\section{Introduction}\label{sec-intro}
De Sitter space plays a central role in cosmology.  In the standard
model, the first moment after the big bang was a period of inflation
in which the universe expanded many times over during an approximately
de Sitter phase.  Recent observations indicate that the expansion is
currently accelerating, indicating the presence of a dark matter
component which is most economically explained by a positive
cosmological constant---in which case the future of our universe is a
de Sitter phase.  Moreover, if string theory is the correct description
of nature, we may expect that our entire observable universe is a
bubble expanding in an eternally inflating false vacuum de Sitter
space \cite{bp, suss1}.  In such a model bubbles of the true vacuum 
nucleate in the false vacuum, and their walls accelerate
outwards under the gravitational pressure due to the differing values
of the vacuum energy inside and out.  Observers may form inside these
cosmic bubbles, which viewed from the inside appear to be negatively
curved expanding Robertson-Walker cosmologies \cite{cdl}. The
cosmology inside will be affected in interesting ways both at its
``Big Bang" \cite{fkrs} and later, by collisions with other bubbles
(see e.g \cite{ckl2}).  If this model is correct, understanding de
Sitter space becomes even more crucial.

But de Sitter space has proved maddeningly difficult to understand.
In eternal de Sitter, points at fixed comoving distance always become
causally disconnected at late times.  As a result, correlation
functions of points separated by more than a single Hubble length are
not observable, and so one cannot define observables (like an S-matrix
or boundary correlation functions in anti-de Sitter space) using
well-separated points in space \cite{witten}.  Moreover, because of
the thermal nature of de Sitter and its finite entropy  one expects
correlators of operators inserted at timelike separated points to be
quasi-periodic functions of the time separation, which neither
converge nor settle into any predictable pattern (and in fact
eventually produce fluctuations to every state consistent with the
conservation laws) \cite{suss3, suss4}.  Attempts to define a dS/CFT
correspondence indicate that if such a dual theory exists, it cannot
be unitary \cite{strom, suss5}.  In the larger ``multiverse" of the
string theory landscape, one would like to understand how to average
over distributions of cosmic bubbles so as to compute the expected
values of cosmological observables visible to those living inside
them.  Much interesting work has been done on this problem;  see e.g \cite{guth} for a review. However, this too has been plagued by infinities and the non-uniqueness in the choice of measure.

\begin{figure}
\begin{center}
\includegraphics[scale=.6]{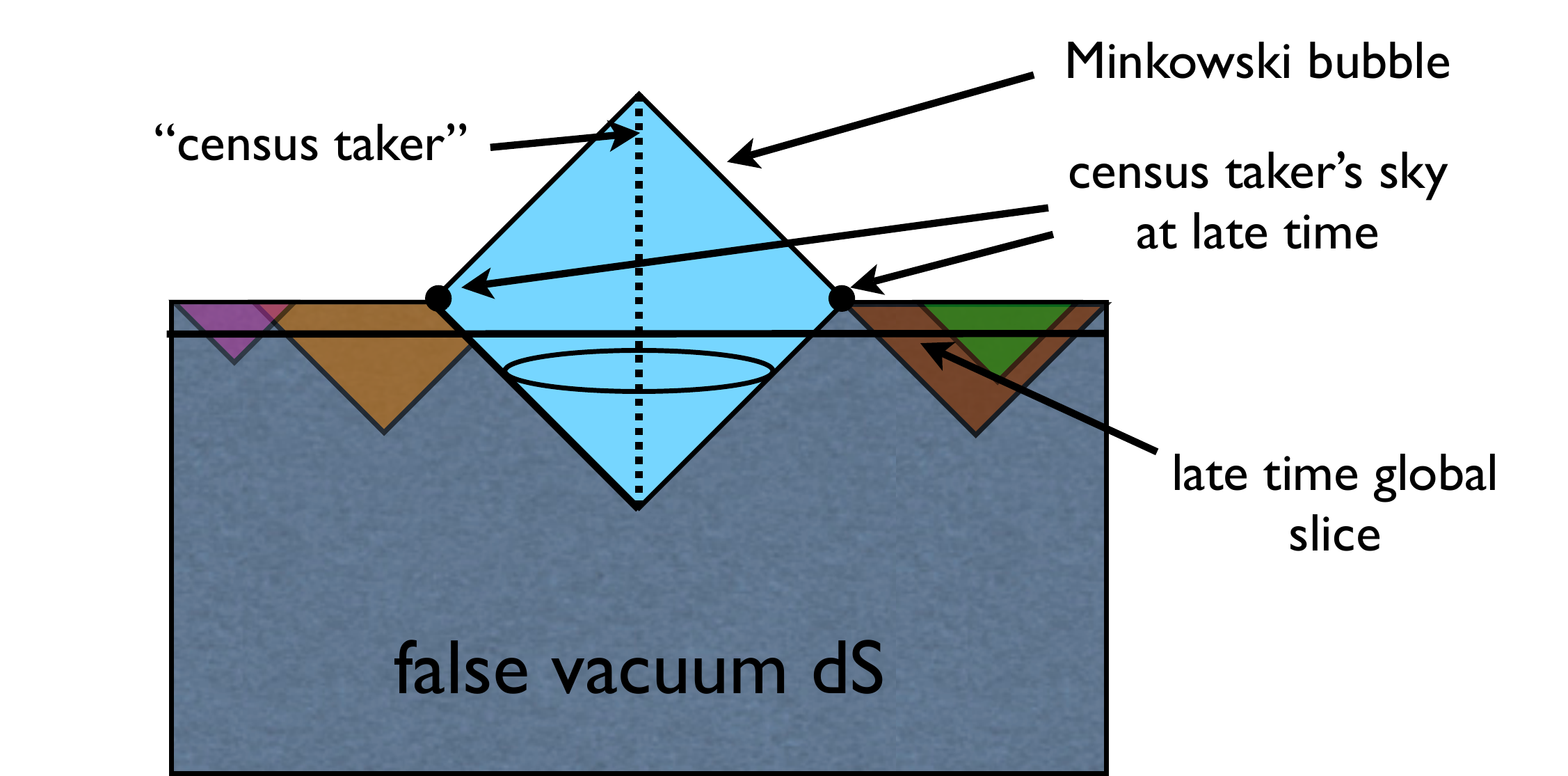}
\end{center}
\caption{A Carter-Penrose conformal diagram for bubbling de Sitter
  space.  One can slice with a constant global time surface, which
  from $d=2+1$ de Sitter would produce a 2-sphere tiled with bubbles
  (see Fig. \ref{globalf}).  Alternatively one can consider the sky as observed at late times by a ``census taker" living in a $\Lambda=0$ bubble.  Starting from $d=3+1$ de Sitter, this will produce a very similar 2-sphere.   }
\label{census}
\end{figure}

\begin{figure}
\begin{center}
\includegraphics[scale=.6]{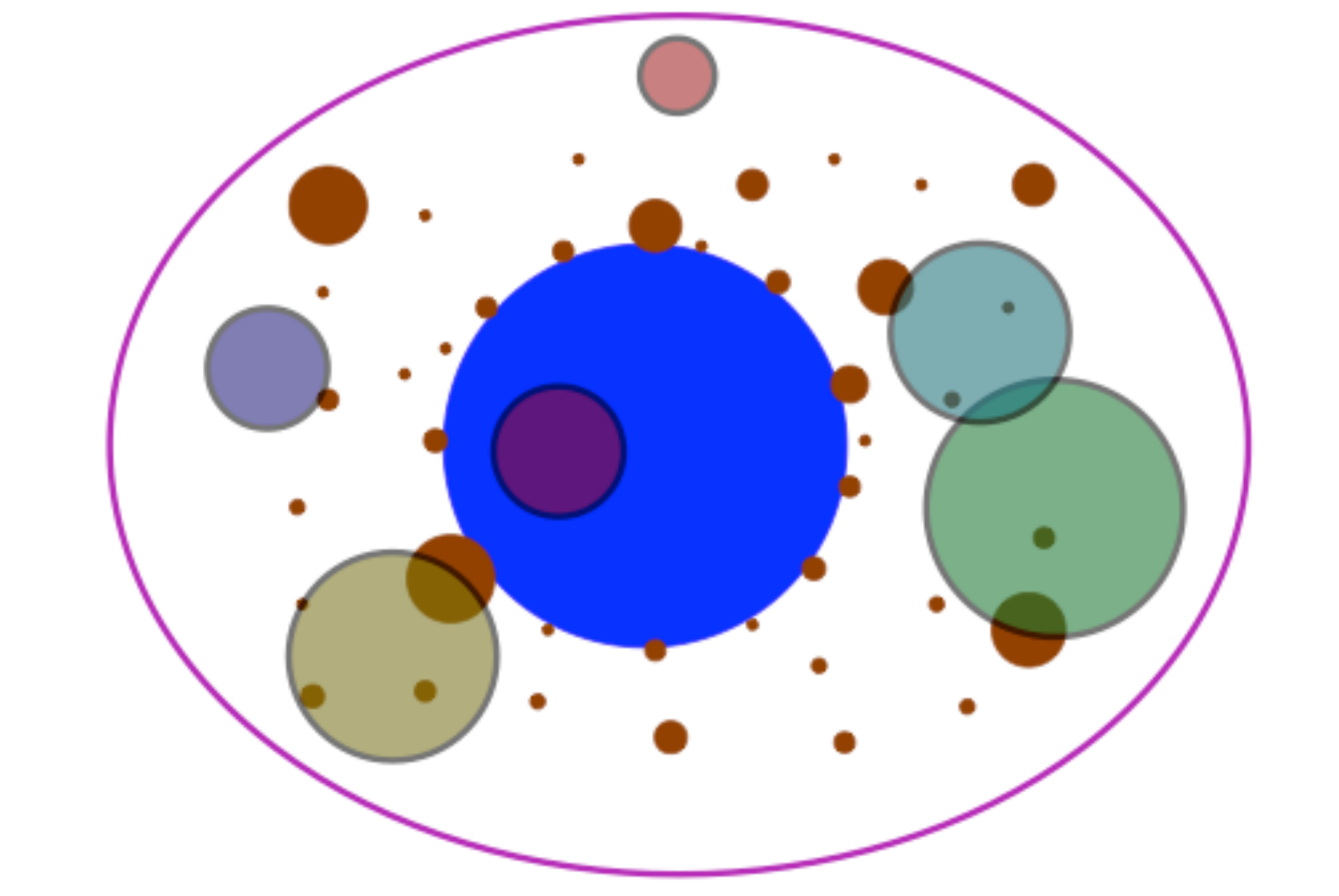}
\end{center}
\caption{Artist's rendition of a global slice of bubbling de Sitter space at late times, or a census taker's sky.}
\label{globalf}
\end{figure}

%In this note we will present a new approach to the problem of defining
%observables in de Sitter. 
Here, we analyze the correlation functions of a putative dual CFT for
eternal inflation.
 Our philosophy is that if one can find a consistent and well-defined
 set of quantities in eternally inflating spaces, this may lead
 to a solution of many of the problems above.
To begin, consider an eternally inflating spacetime in $d+1$ spacetime dimensions in which at least one interacting scalar field is undergoing tunneling and forming bubbles.  If one takes a $d$-dimensional constant time slice across
the spacetime at late time, the slice will contain many casually
disconnected regions and many vacuum bubbles, as shown in figures
\ref{census} and \ref{globalf}.
  Each bubble
appears initially with small size, but as time passes its radius
grows, asymptoting to a finite comoving radius determined by the
conformal time at which it appeared (late appearing bubbles are
smaller).  

We make some simplifying assumptions which allow us to compute
the statistical distribution of these bubbles on the $d$-slice. (For a
more general discussion, see \cite{gv}.) This approach has the advantage
of simplicity, but the disadvantage that these bubble distributions are not observable.

Any given observer can observe only a subset of the eternally
inflating spacetime. Consider an observer inside a bubble of some
type, the ``observation bubble.'' The observation bubble collides with
other bubbles
that form through quantum nucleation in the false vacuum nearby, as
shown in figure \ref{census}.  Each
collision bubble will affect the observation bubble's wall inside a
ball (a disk in the case of an observer 3+1 dimensional de Sitter).
As time passes for an observer inside the observation bubble the collision
appears as a point and grows at a rate which asymptotically approaches
that of the observation bubble's wall itself.  As a result the
angular radius of the disk asymptotes at late time to some finite size $\psi$,
which is determined in a simple way by the time of the collision (late
collisions make smaller disks).  If the observer has access to large
conformal times---which requires that the cosmological constant inside
the observation bubble be small---she can observe this asymptotic
distribution by, for example, its effects on her cosmic microwave
background sky.  Such an observer has been referred to as a ``census
taker" \cite{census}.

We will investigate the statistics of the distribution of these
bubbles and their collisions, focusing primarily on the case of 2D
distributions. The distribution can be thought of as describing
either the set of bubbles on a late time slice in 2+1
dimensional de Sitter space, or the set of collisions with an
observation bubble in 3+1 de Sitter.
 As we will see, using the bubble distribution on these
surfaces one can define a set of correlation functions which behave
like primary operators of a 2D conformal field theory.  The operators
are the exponentials of a discrete number operator that counts the
number of disks that overlap the point where it is inserted.  

Using
the distribution we compute the 2--,3--, and 4-point functions of
these exponential operators analytically and exactly.  These turn out
to be consistent with the hypothesis that these exponential operators
are primary fields of a conformal field theory, except that the
4-point function exhibits a non-analyticity as a function of the
location of the operator insertions.  The theory is conformally
invariant for arbitrary values of the dimensionless tunneling rate
$\gamma$.
The relation between the charge
$\beta$ of the exponential of the number operator $e^{i \beta N(z)}$
and its scaling dimension is interesting and novel: \be \Delta \left(
  \beta \right) = \pi \gamma (1-\cos \beta).  \ee We compute the
central charge of this putative conformal field theory by evaluating
its partition function on a sphere of radius $R$, and find a result
proportional to the continuous parameter $\gamma$.

We also discuss higher dimensional versions of the theory.  The
simplest such extension is a 3D version obtained from the statistics
of bubbles on 3D global slices of $3+1$ de Sitter.  This 3D theory has
a dimension zero number operator, and its exponentials again behave
like primary fields with positive dimension.  In fact, we can show
that in any dimension $d$, the theory always contains a dimension zero
number operator and exponential operators constructed from it with
arbitrarily small positive weight.  Since no such operator can exist in a
unitary field theory in more than $d=2$, these higher dimensional
theories cannot be unitary.

Finally, we will demonstrate that in the limit that the decay rate $\gamma \to \infty$, the theory in any number of dimensions becomes free:  correlators of the exponential operators factorize onto products of 2-point functions in precisely the same way as vertex operators for a free, massless scalar in two dimensions.

\section{Conformal invariance}

The simplest case to consider is one in which the cosmological
constant inside the bubbles is the same as that of the ``false vacuum"
outside, and where bubbles can nucleate both inside and outside other
bubbles---always with the same decay rate $\gamma$, where $\gamma$ is
defined as the dimensionless decay rate per unit Hubble time per unit
Hubble volume. 

\begin{figure}
\label{pot}
\begin{center}
\includegraphics[scale=.4]{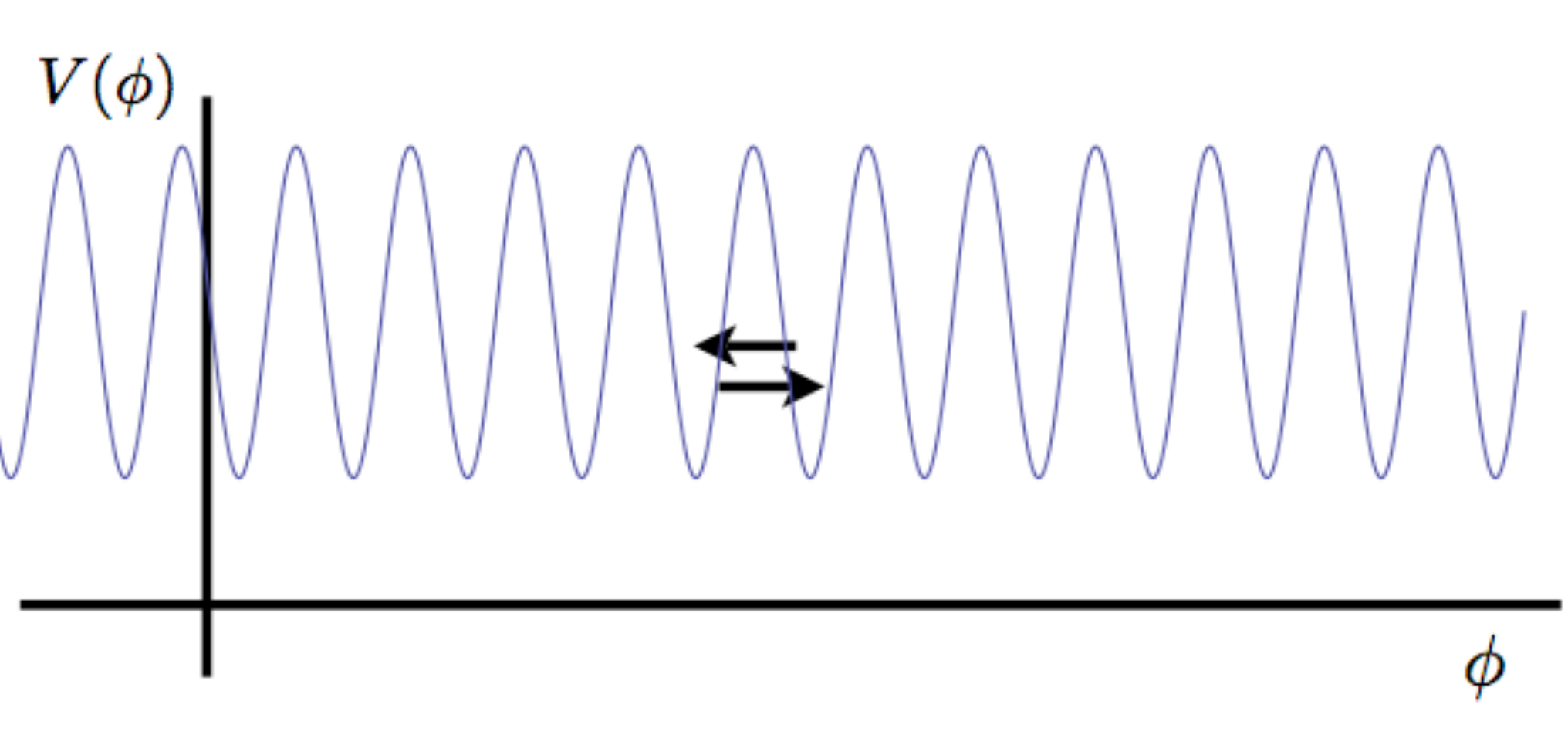}
\end{center}
\caption{A potential for a scalar field coupled to gravity that could produce bubble distributions of the type we consider.}
\label{potential}
\end{figure}

In order for the tunneling rate to be constant, all of the vacua
should be identical. The simplest model is the potential drawn in
figure \ref{potential}, where the potential has a discrete shift symmetry relating
the vacua to each other. 
Even once the potential is symmetric, there
are in general nontrivial interactions between bubbles, such as
correlations between their nucleation points and collisions between bubbles. We work
in the noninteracting limit where the nucleation rate is constant,
independent of the presence of other bubbles.

With these assumptions one can compute the bubble
distribution on a global time slice.  The spacetime is de Sitter with
metric 
\be ds^2 = { 1 \over \sin^2 \eta} \left( -d \eta^2 +
  d\Omega_{d}^2 \right) \ .  
\ee 
where $d\Omega_d^2$ is the metric on a $d$-sphere.
The number of bubbles that nucleate
in a conformal time interval $d \eta$ is proportional to the spacetime
volume available, 
\be  dN = \gamma dV_{d+1} = \gamma {d
  \eta \over \sin^{d+1} \eta} d\Omega_{d} 
\ee 
where $dV_d$ is the
$d-$dimensional spacetime volume element and the $d\Omega_{d}$ factor
refers to the location of the nucleation point on the spatial slice.

\begin{figure}
\begin{center}
\includegraphics[scale=.6, angle=-90]{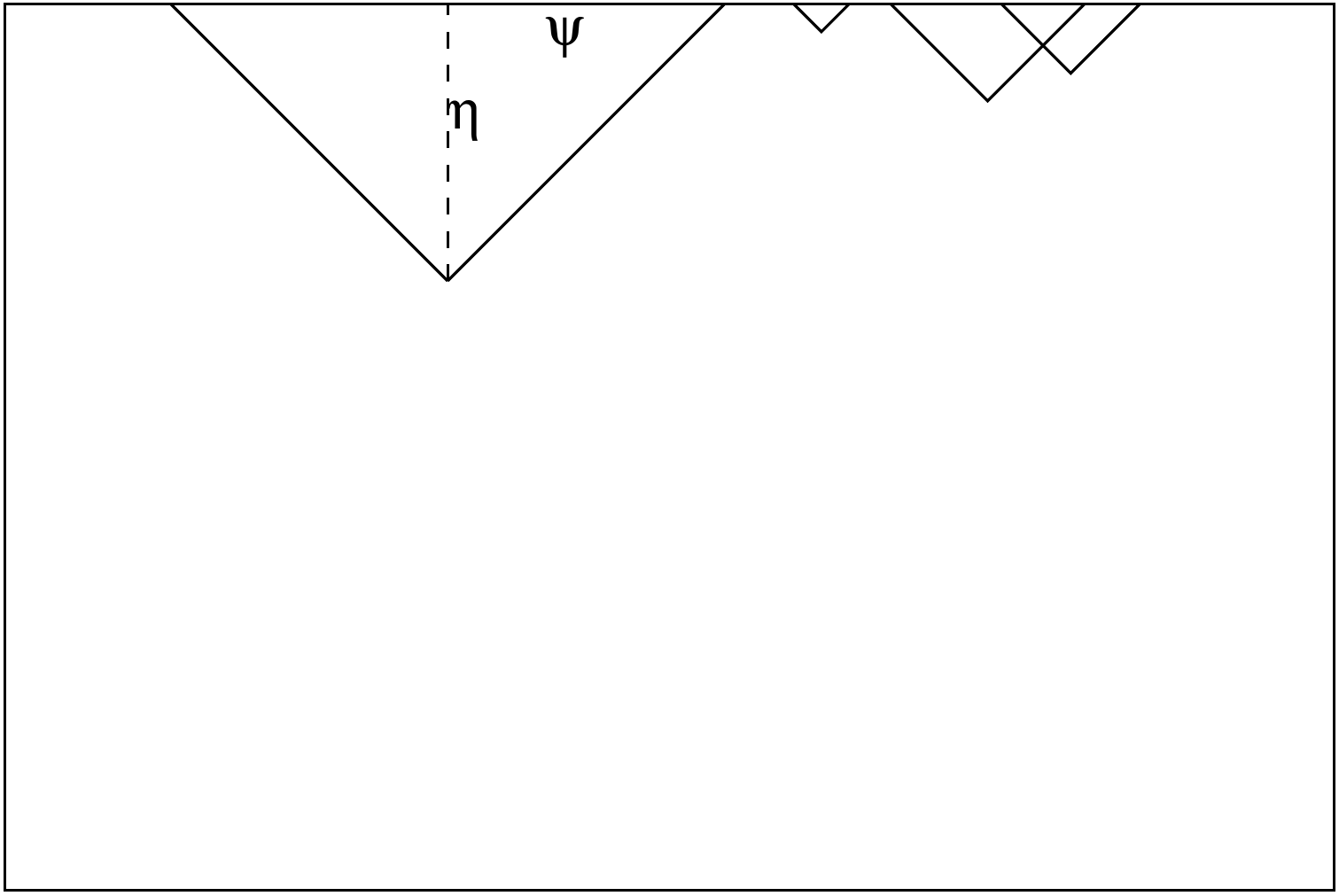}
\end{center}
\caption{A bubble nucleated at conformal time $\eta$ covers an angular
  size $\psi = |\eta|$ on future infinity.}
\label{etapsi}
\end{figure}

We want to characterize the distribution of bubbles on future infinity
of de Sitter space, which is given by $\eta = 0$ in these
coordinates. A given bubble nucleation nucleation will affect a ball on future infinity.
The domain wall of a bubble asymptotically approaches the future lightcone of the nucleation point. Light rays satisfy
\be
d \psi  =  d \eta
\ee
where $\psi$ is the angular radius of the lightcone. Therefore, a
bubble nucleated at time $\eta$ has angular size $\psi = |\eta|$ on
the conformal boundary, as shown in figure \ref{etapsi}.
 The distribution of bubbles on the boundary is then
\be
dN = \gamma {d \psi \over \sin^{d+1} \psi} d\Omega_{d} \ .
\label{gdist}
\ee
Previous attempts to define a theory on such slices can be found in \cite{strom, gv}. 

As mentioned in the introduction, we can also consider the
distribution of bubbles which collide with a given ``observation
bubble.''
In \cite{fkns} we computed this distribution: the distribution of collision bubbles on
the boundary of the future lightcone of a point in 3+1 de Sitter space
(i.e., the observation bubble's wall) after infinite time.  The number
distribution takes the following simple form:

\begin{figure}
\begin{center}
\subfigure[]{\includegraphics[scale=.35]{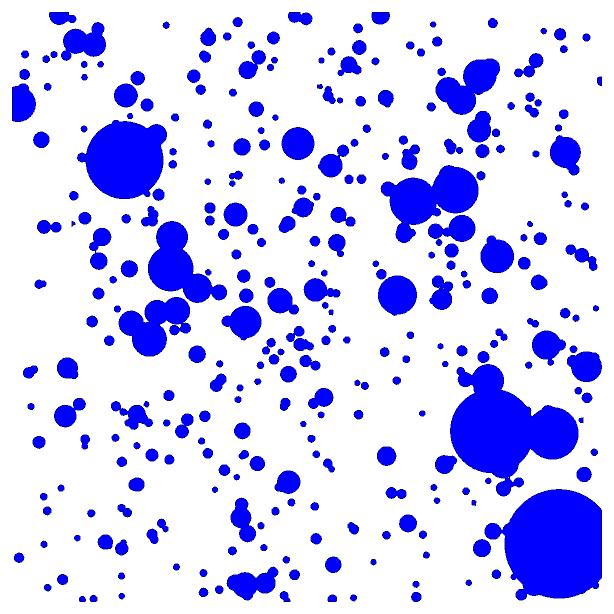}}
\hspace{1cm}
\subfigure[]{\includegraphics[scale=.35]{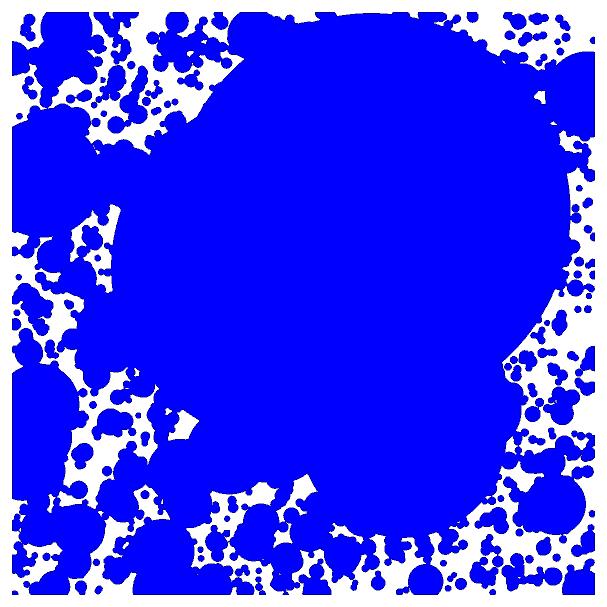}}
\end{center}
\caption{Numerical simulations of the model in $d=2$.  Only one disk type is shown.  Left pane:  $\gamma = .1, \delta/R=.01$.  Right pane:  $\gamma = .5, \delta/R=.01$.}
\label{numerics}
\end{figure}
\be \label{sdist}
dN(\psi, \theta, \phi) = {4 \gamma \over 3} {d\psi \over \sin^3 \psi} \sin \theta d \theta d \phi,
\ee
where 
%$\gamma$ is the decay rate per unit time per unit volume in units of the false vacuum Hubble length, 
$\psi \in (0, \pi)$ is again the angular radius of the disk, and $\theta$ and $\phi$ are the coordinates of its center on the boundary sphere (we have set the radius of the sphere to 1 for convenience).  
% We will allow overlaps without any change - in other words, we will allow bubbles to nucleate inside other bubbles with a decay rate identical to that of the parent false vacuum. XXXX MAKE GENERAL DIMENSION? XXX  
Notice that this distribution is identical in form to Eq. \ref{gdist} with $d=2$.  An attempt to define a conformal field theory on this sphere can be found in \cite{fssy}. More generally, the number distribution of collisions on the boundary of an ``observation bubble" in $d+2$ dimensional de Sitter space is
\be \label{sd2}
dN = \gamma {\Omega_{d +2} \over \Omega_{d+1}} {d\psi \over \sin^{d+1} \psi} d \Omega_{d}
\ee
where $\Omega_n$ is the surface area of a unit $n$-sphere.

This distribution turns out to have the remarkable property that it is
invariant under global conformal transformations $SO(d+1,1)$, which
are Mobius transformations in $d=2$.  The easiest way to see this is to
stereographically project the distribution to the plane.  This
projection maps spheres to spheres. In terms of the coordinates
$x^i$ of the center of the sphere and the sphere radius $r$, a little algebra shows that the distribution becomes simply
\be \label{pdist}
dN  = \gamma {dr \over r^{d+1}} { d^d x },
\ee
where $x^i$ are the Cartesian coordinates on the plane.
% where $z = (x+iy)/\sqrt{2}$ and $d^2 z = i d z d\bar z$. 
 Note that this is also the small angular radius, small area approximation to the sphere distribution (\ref{gdist}).  Global conformal transformations on the sphere are generated by rotations plus special conformal transformations.  In the stereographic plane, special conformal transformations around the origin take a particularly simple form---they are simply the scalings $x^i \rightarrow \lambda x^i$ and $r \rightarrow \lambda r$.  Therefore the distribution (\ref{gdist}) is conformally invariant, since it is manifestly rotation invariant, and one can always choose the origin of the stereographic plane to coincide with one of the fixed points of the special conformal transformation.  As noted in \cite{fkns}, this symmetry group is the set of Lorentz transformations on the point of nucleation of the observation bubble.

\subsection{Fractals and percolation}

Benoit Mandelbrot considered the distribution Eq. \ref{pdist} in
\cite{mandelbrot}, where he commented that in $d=2$ it approximates
the distribution of craters on the moon.  One can easily calculate the
Minkowski or ``box-counting" fractal dimension of various sets of
points defined by this distribution.  For example, the set of points
not inside any bubble is a set of measure zero (because the set of
spheres with radii in any given logarithmic interval covers a finite
fraction of the volume).  To compute the fractal dimension of this
set, one chooses boxes of linear dimension $\epsilon$, finds the
minimum number $N_b(\epsilon)$ of boxes necessary to fully cover the
set, and then the dimension is defined as $d_F = \lim_{\epsilon \to 0}
(-\partial \ln N_b /\partial \ln \epsilon)$.

To compute this, note that the volume $V(\epsilon)$ which remains uncovered by spheres of radius greater than $\epsilon$ is the volume $V(\epsilon + d\epsilon)$ minus the volume covered by spheres of size between $\epsilon$ and $\epsilon + d\epsilon$:
\be
dV(\epsilon) = -C_d \epsilon^d dN
\ee
 where $C_d$ is the volume of a unit $d$-sphere. The infinitesimal number of spheres is given by
 \be
 dN = \gamma {d \epsilon \over \epsilon^{d+1}} V(\epsilon) \ ,
 \ee
 so we have the differential equation
 \be
 {dV \over V} = - \gamma C_d {d \epsilon \over \epsilon}
 \ee 
  Integrating this equation gives $V(\epsilon) = V_0 \epsilon^{\gamma C_d}$.  
The number of cubical boxes of size $\epsilon$ required to cover this area is simply $N_b(\epsilon) = V(\epsilon) \epsilon^{-d}$, and therefore    \cite{mandelbrot} 
 \be
  d_F = d - \gamma C_d
  \ee
(see \cite{vilenkin} for this calculation in the context of bubbles in $3+1$ de Sitter).
Similarly one could compute the dimension of other sets, such as the set of points covered by exactly, or at most, $k$ spheres.  These will be scale invariant fractal sets of measure zero as well.  The existence of these fractals is not surprising given the statistical scale invariance of the distribution.  
%However the distribution is more than merely scale invariant---it is invariant under the full global conformal group, which raises the possibility that one may be able to define a conformal field theory from its correlation functions.

One may expect that these sets will undergo percolation transitions at special values of $\gamma$ \cite{mandelbrot}.  For example in $D=2$, as one increases $\gamma$ from zero there should be a percolation critical point where the set transitions from connected along filaments to a disconnected dust.  This will occur at some $d_F \sim 1$.  In $d=3$ one expects two such transitions:  from ``ramified veils" to filaments at $d_F \sim 2$, and from filaments to dust at some other $d_F \sim 1$ \cite{mandelbrot}.

If in fact one can regard this as a type of percolation model, it has some novel properties.  The most striking is that as we will see, the theory seems to be conformally invariant for all values of $\gamma$ rather than just at the critical points---or at least one can define operators that transform covariantly for all $\gamma$.  Additionally, at least some correlation functions can be computed analytically and exactly up to at least the 4-point function.  It would be interesting to see if quantities of primary interest for percolation (for example crossing probabilities) could be extracted using the techniques developed here, but we leave this question for future work.

Models of a somewhat similar type have been considered in the past under the name ``continuum fractal percolation," where ``continuum" refers to the lack of an underlying lattice, and ``fractal" to some self-similarity in the distribution of disks (or other shapes---see e.g. \cite{meester}).  Another model with some similarities is ``Mandelbrot percolation," in which a square is subdivided into $N^2$ smaller squares for some integer $N$, each of which is colored with a probability $p$, and then the remaining uncolored squares are subdivided and the process repeats.  Interestingly, while this model does have a percolation transition, it is first order \cite{chayes}---and hence not conformal even at the transition!  Percolation in power-law disk distributions was considered in the context of networks in \cite{wss}.

\section{Correlation functions} \label{sec-col}

In this note we will concentrate on the correlation functions of
a field with multiple vacua, as pictured in figure \ref{potential}. We label the minima by $N$, where $N$ takes integer values. As mentioned earlier, we assume that the vacua are identical, so there is a shift symmetry $N \to N + 1$. Starting from a given vacuum, the field can tunnel to the left or to the right; we call these events instantons and anti-instantons.

When the forward lightcones of two spacelike separated nucleation points overlap, additional physics is needed to determine the field profile in the overlap region. If the critical bubble size is small, one can think of these overlap regions as collisions between bubbles. It is then nontrivial and model dependent to solve for what happens in the future of a collision. In our model where the vacua are degenerate, the critical bubble is horizon size, so instead of collisions it is more accurate to think of bubbles nucleating on top of existing domain walls. We make the simplest possible assumption about the overlap regions: we assume the instantons satisfy superposition. In a region to the future of the nucleation points of $N_+$ instantons and $N_-$ anti-instantons, we assume the field is in the minimum $N = N_+ - N_-$.  
These simple assumptions allow us to calculate correlation functions explicitly, but it would clearly be interesting to perturb away from them by allowing interactions between nucleation points and nontrivial dynamics in the overlap regions. For small tunneling rate $\gamma$, the instantons are very dilute and one may expect that the interactions are unimportant.

It is convenient to construct a partition function for the bubble distribution with which to compute expectation values.  For convenience we will work on the plane, using the distribution (\ref{pdist}).  The partition function is
\be \label{ppart}
Z = \sum_{n=0}^\infty {\gamma^n \over n!} \prod_{k=1}^n \left(p_+ +
  p_- \right)^n \int_\delta^R {d r_k \over r_k^3} \int d^2x_k  = \exp{
  \( \gamma (p_+  + p_-) \int_\delta^R {d r \over r^3} \int d^2z  \)} .
\ee
Each term in the sum corresponds to a configuration of $n$ disks, with
the $k$th disk centered at the point $x_k$ and with radius $r_k$. The
factor $(p_+ + p_-)$ denotes the probabilities of the
two possible types of nucleations,
left-moving and right-moving. Because all of the vacua are identical
by assumption, detailed balance demands $p_+ = p_- = 1/2$.
  To avoid infinities the integral must be cut off in both small and large disk sizes, although as we will see, well-defined correlation functions on the plane do not depend on the IR cutoff $R$.  The factor of $\gamma^n$ is the appropriate weight for a configuration of $n$ disks, given that $\gamma \sim e^{-S_{\rm inst}}$ and that instanton interactions can be neglected.  %Finally, the factor of $1/n!$ arises because we are treating the disks as indistinguishable, and the reason for the subscript on $Z_+$ will become clear shortly.
The partition function factorizes into instanton and anti-instanton pieces,
\be
Z = Z_+ Z_-
\ee
with 
\be
Z_+ =  \exp{ \( {\gamma_+} \int_\delta^R {d r \over r^3} \int d^2z  \)} \ .
\ee
Symmetry determines $\gamma_+ = \gamma_- = \gamma/2$.

Because the distribution is invariant under Mobius transformations, one expects correlation functions of well-defined operators computed using the partition function \ref{ppart} to be Mobius covariant.  The subtlety arises from the cutoffs---divergent correlators will not transform simply under Mobius transformations, but as we will see
one can define well-behaved operators with correlation functons that transform simply.

The potential has a discrete shift symmetry $N \to N + 1$. The natural operators to consider have definite charge under the shift symmetry.   The simplest such operators are exponentials,
\be
V_\beta (z) \equiv e^{i \beta N(z)} \ .
\ee
%
%One way to remove the IR divergence in the correlators of massless fields is to take their exponentials.  As we will see, the operators $V_\beta(z) \equiv e^{i \sqrt{2} \alpha \phi(z)}= e^{  i \beta \left( N_+(z) - N_-(z) \right) }$, where $\beta = \alpha \sqrt{2 \over \pi \gamma}$, 
We will see that these operators have positive definite weight at
least under Mobius transformations, and their correlators are finite
in the IR. Correlators of $N$ itself can be determined by
differentiating the correlators of exponentials.  As we will see this
gives logarithms, as one would expect if $N$ were a massless field
with dimension zero.  Since the correlation functions of such fields
are not well-defined, this is another reason to consider exponentials.

\subsection{The 1-point function}

To compute the 1-point function $\la V_\beta(z) \ra$  one simply needs to insert it into the partition sum (\ref{ppart}):
\be
\la V_\beta(z) \ra = Z^{-1} \left( \sum_{n=0}^\infty {\gamma_+^n
  \over n!} \prod_{k=1}^n \int_\delta^R {d r_k \over r_k^3} 
  \int d^2 x_k e^{i \beta  N_+(z)          }\right) 
\left( \sum_{n=0}^\infty {\gamma_-^n
  \over n!} \prod_{k=1}^n \int_\delta^R {d r_k \over r_k^3} \int d^2 x_k e^{- i \beta  N_-(z))}\right)
\ee
where the two terms correspond to the instantons and
anti-instantons. The exponential operator has a simple product
form, and the contributions from the instantons and anti-instantons
are complex conjugates of each other, so
\be
\la V_\beta(z_1) \ra = Z^{-1} \left| \exp \left\{ \gamma_+ \int_\delta^R {d r \over r^3} \int d^2x
 \left[ e^{i \beta} \Theta(2 r - |x - z_1|) + \Theta(- 2 r +
|x - z_1|) \right] \right\} \right|^2
\ee
Cancelling against $Z^{-1}$ and collecting terms, this becomes
\be
\la V_\beta(z_1) \ra =  
\exp \left[ -  \gamma (1 - \cos \beta)  \int_\delta^R {d r \over r^3} \int d^2x
  \Theta(2 r - |x - z_1|) \right]
\ee
The integral over $x$ gives the area $A_1(r)$ such that a disk of radius $r$ covers
$z_1$ if its center is contained in $A_1$. This set of points is a
disk of radius $r$ centered at $z$.  Then
\be
\la V_\beta(z_1) \ra =  
\exp \left[ -  \gamma (1 - \cos \beta) \int_\delta^R {d r \over r^3}
A_1(r) \right]
\ee
The area is $A_1 = \pi r^2$, so the integral is
\be
I_1 \equiv {1 \over \pi} \int_\delta^R {d r \over r^3}
A_1(r) =  \ln {R \over \delta}
\ee
%% \be
%% \la V_\beta(z) \ra = Z_+^{-2}  e^{ \gamma I(\alpha)}e^{\gamma I(-\alpha)} = e^{\gamma \left( -2I(0) + I(\beta) + I(-\beta) \right) },
%% \ee 
%% where $\beta \equiv \alpha/\sqrt{\pi \gamma}$, $I(\beta) =  \int_\delta^R {d r_k \over r_k^3} \int dx_k d \bar x_k e^{i \beta N(z)}$ and the $-2 I(0)$ term come from $Z_+^{-2}$.  A quick computation gives 
%% \be
%% I(\beta) - I(0) = \left( e^{i \beta}-1 \right) \int_\delta^R \pi r^2 {dr \over r^3} =  
%%  \left( e^{i \beta}-1 \right) \ln {R \over \delta},
%%  \ee
%% and so
Finally, the one point function of the exponential operator is
\be
\la V_\beta(z) \ra = \left( {R \over \delta} \right)^{- \pi \gamma (1
  -  \cos \beta)} = \left\{
  \begin{array}{ c c }
     1 & {\rm if ~ \beta =} 2 \pi n ,~n\in {\cal Z}\\
     0 & {\rm otherwise}
  \end{array} 
\right\}.
\ee
This type of ``conservation of charge" condition is familiar from Liouville theory and free scalar CFTs, but the periodicity in $\beta$ (which is a consequence of the quantization of $N$) is novel.

\subsection{The 2-point function}

One can compute the 2-point function by the same techniques.  A
similar analysis to the one above gives

\beq\label{2point}
& \la V_{\beta_1}(z_1)  V_{\beta_2}(z_2) \ra = \left| \la \exp \left[
  i \beta_1 N(z_1) + i \beta_2 N(z_2)  \right] \ra \right|^2   = \\ & \exp{\left\{ -\gamma  \int {d r \over r^3} \int d^2x    \left(1 -  \cos \left[ \beta_1 \Theta(r- |x-z_1|)+\beta_2 \Theta(r- |x-z_2|)   \right] \right) \right\}. } 
\eeq
This is equivalent to
\be
\la V_{\beta_1}(z_1)  V_{\beta_2}(z_2) \ra =
\exp \left\{ - \gamma \int {dr \over r^3} \left[ (1 - \cos
  \beta_1) \ea_1 + (1 - \cos \beta_2) \ea_2 + (1 - \cos(\beta_1 +
  \beta_2)) \ea_{12} \right]\right\}
\ee
where we have defined the ``exclusive area'' $\ea_1 $ as the area of
the region such that disks centered in that region cover $z_1$ but not
$z_2$. Equivalently, construct two disks of radius $r$, one centered
at $z_1$ and the other centered at $z_2$. Then $\ea_1$ is the area
covered $only$ by the disk centered at $z_1$.

Similarly, we define the integral over these areas as
\be\label{eq2}
 \ev_{1}(z_1, z_2) \equiv {1 \over \pi} \int {d r \over r^3} \ea_1 (r, z_1, z_2)~.
 \ee
With this notation, the two point function is
\be
\la V_{\beta_1}(z_1)  V_{\beta_2}(z_2) \ra =
\exp \left\{-  \pi \gamma \left[ (1 - \cos \beta_1) \ev_1  + (1 - \cos \beta_2) \ev_2 + (1 - \cos (\beta_1 + \beta_2))
  \ev_{12}  \right] \right\}
\label{eqi}
\ee
The integrals $\ev_{ij...}$ have a simple
interpretation as spacetime volumes in de Sitter space: $\ev_{ijk}$ is
proportional to the spacetime volume available to nucleate bubbles
which cover $only$ the points $(z_i, z_j, z_k)$.

To evaluate the integrals, we need to write the exclusive areas in
terms of the simpler inclusive areas. For the two-point function, we
have
\begin{eqnarray}
\ea_1 &=& A_1 - A_{12} \\
\ea_2 &=& A_2 - A_{12} \\
\ea_{12} &=& A_{12}
\end{eqnarray}
The same equations hold for the integrals of the areas,
\begin{eqnarray}
\ev_1 &=& \nev_1 - \nev_{12} \\
\ev_2 &=& \nev_2 - \nev_{12} \\
\ev_{12} &=& \nev_{12}
\end{eqnarray}
We need to compute $I_{12}$, the integral of the area covered by both
disks. Some simple geometry yields
\be
A_{12}(r, z_{1}, z_2) = 2 r^2 \left( \cos^{-1}(|z_{12}|/2r) - (|z_{12}|/2r)\sqrt{1-(|z_{12}|/2r)^2} \right) \theta(2 r-|z_{12}|),
\ee
where $|z_{12}| \equiv |z_1-z_2|$ is the distance between the centers of the disks.  Integrating this gives
\be
I_{12}(r, z_1, z_2) = {1 \over \pi} \int_\delta^R A_{12} ~dr/r^3 =  \ln(R/|z_{12}|) - 1/2 + {\cal O}(1/R).
\ee
Plugging this in gives
\begin{eqnarray}
\ev_1 = \ev_2 =  \ln {|z_{12}| \over \delta} + {1 \over 2} \\
\ev_{12} =  \ln {R \over |z_{12}|} - {1 \over 2}
\end{eqnarray}
Let us redefine the UV cutoff $\delta$ to eliminate the annoying
constant factor so that
\begin{eqnarray}
\ev_1 = \ev_2 =  \ln {|z_{12}| \over \delta} 
\end{eqnarray}

 %%  The integral splits up into four areas SEE FIGURE XZXX, labeled $A_!, A_2, A_{12}$, and the region outside both disks, so that
%% \be\label{I12}
%% I_{12} = \int {d r \over r^3}\left( e^{i \beta_1}(A_1-A_{12}) + e^{i \beta_2}(A_2-A_{12}) + e^{i(\beta_1+\beta_2)}A_{12} - A_1 - A_2 + A_{12} \right),
%% \ee
%% where the last three terms come from the -1 in Eq. \ref{eq2}.
Note that $\ev_1$ and $\ev_2$ are infrared finite, while $\ev_{12}$
diverges as $R \to \infty$. This corresponds to an infinite
expected number of disks covering both points 1 and 2. 
Therefore, for $\beta_1 \neq \beta_2$,  the exponent of Eq. \ref{eqi} goes to
$-\infty$ as the infrared cutoff $R$ is taken to infinity. So
the two-point function Eq. \ref{2point} is zero due to IR divergences.
To cancel this divergence it is necessary and sufficient to require
that the coefficient of the double overlap region $\ev_{12}$ is zero;
in other words one needs $\cos(\beta_1+\beta_2)=1$, or  $\beta_1 +
\beta_2 = 2 \pi n$ ($n$ an integer). 

This condition is a kind of charge conservation condition: under the
shift symmetry $N \to N+1$, the operator $V_\beta$ transforms as
$V_\beta \to \exp(i \beta) V_\beta$. So the correlators are nonzero
only when they are invariant under the shift symmetry. Because $N$
takes integer values, the operator $
e^{i \beta N(z)} $ is equivalent to the operator $ e^{i (\beta + 2
  \pi) N(z)}$, so it is natural that the charge conservation condition
is defined mod $2 \pi$. We will see in a moment that the dimensions of
operators are also  periodic functions of $\beta$.

 Enforcing this condition we get
%% \be
%% I_{12} = \left( \cos \beta_1 - 1 \right) \left( 2 \pi \ln {|z_{12}| \over \delta} - \pi +... \right).
%% \ee
%% The $-\pi$ can be absorbed by redefining $\delta$, which we will do, (THIS COULD ACTUALLY MATTER - IF YOU GET A DIFFERENT CONSTANT IN THE 3 OR 4-POINT FUNCTIONS, IT MATTERS.  BUT I DON'T THINK YOU DO)
%% and so in the end we obtain
\be
 \la V_{\beta_1}(z_1)  V_{\beta_2}(z_2) \ra = \left( {\delta \over z_1
     - z_2} \right)^{ \displaystyle \pi \gamma \left(1-\cos \beta_1 \right)}  \left( {\delta \over \bar z_1 - \bar z_2} \right)^{\displaystyle \pi \gamma \left(1-\cos \beta_1 \right)}
 \ee
when $\beta_1 + \beta_2 = 2 \pi n $ (else the correlator is zero).
Defining $\Delta(\beta) = \bar \Delta(\beta) =  { \pi \over 2} \gamma \left(1-\cos \beta \right)$,
this has the form of a two-point function for a conformal operator of dimension $(\Delta, \bar \Delta)$.

\subsubsection{Correlation functions of $N$}
At this point, we pause for a moment in our analysis of exponential
operators to consider quantities which may seem more basic: 
correlators of the field $N$ itself. Because $N \to - N$ is a
symmetry, the 1-point function vanishes, $\langle N(z) \rangle =
0$. The 2-point function can be obtained by differentiating the
2-point function of exponentials,
\be
\langle N(z_1) N(z_2) \rangle = - {\partial \over \partial
  \beta_1}{\partial \over \partial \beta_2}  \la V_{\beta_1}(z_1)
V_{\beta_2}(z_2) \ra 
\ee
evaluated at $\beta_1 = \beta_2 = 0$. A convenient form of the
correlator to differentiate is (\ref{eqi}). Differentiating and
setting $\beta_1 = \beta_2 = 0$ gives
\be
\langle N(z_1) N(z_2) \rangle = \pi \gamma \ev_{12} = \pi \gamma
\left( \ln {R \over
  \left| z_{12} \right| } - {1 \over 2} \right)
\ee
This is the correlation function of a dimension zero field. To put it
in a more standard form, the
additive factor of $1/2$ could be absorbed into the infrared cutoff
and the prefactor $\pi \gamma$ could be absorbed into a field
redefinition of $N$. The
presence of the infrared divergence means that these correlators are
not really well-defined. This is not surprising: since the theory has
a symmetry $N \to N+1$, $N$ is not a physical
quantity. However, the exponentials we have been considering are
physical and have well-defined correlators. Exactly the same issues
arise for a free massless scalar in two dimensions.

\subsection{The 3-point function}

To compute the 3-point function of exponentials $ \la V_{\beta_1}(z_1)  V_{\beta_2}(z_2) V_{\beta_3}(z_3) \ra$ we will need to evaluate the integrals of the overlap regions of three disks of equal size, each centered on a point $z_i$ where the operators is inserted.  The calculation proceeds along the same lines as for the 2-point function;
omitting some details one obtains % (c.f. Eq. \ref{I12}) 
\begin{eqnarray*}
 \la V_{\beta_1}(z_1) V_{\beta_2}(z_2) V_{\beta_3}(z_3) \ra = 
\exp \left\{ -  \pi \gamma \left[ (1 - \cos \beta_1) \ev_1 + (1 - \cos
  \beta_2) \ev_2 +(1 - \cos \beta_3) \ev_3 \right. \right. +& \\ 
 (1 - \cos (\beta_1 + \beta_2)) \ev_{12} +(1 - \cos (\beta_1 + \beta_3)) \ev_{13} +(1 - \cos (\beta_2 + \beta_3)) \ev_{23}  +
\left. \left. (1 - \cos (\beta_1 + \beta_2 + \beta_3)) \ev_{123} \right] \right\}&.
\end{eqnarray*}

%% \beq
%% \int {dr \over r^3} \left\{ e^{i \beta_1} \left(A_1 - A_{12} - A_{13} + A_{123} \right) + e^{i(\beta_1 + \beta_2)} \left( A_{12} - A_{123} \right) + e^{i(\beta_1+\beta_2+\beta_3)}A_{123} \right. \\ \left.  - \left(A_1- A_{12}+A_{123} \right)
%%    + {\rm cyclic~permutations} \right\}.
%% \eeq

Since the integral  $\ev_{123}$ is again logarithmically divergent
at large $r$, a  ``charge cancellation" condition is required to
cancel the IR divergence that would otherwise send the correlator to
zero. 
 %% The relevant term is
%% \be
%%  \left( e^{i \beta_1} + e^{i \beta_2} +e^{i \beta_3} -e^{i (\beta_1+ \beta_2)}  -e^{i (\beta_2+ \beta_3)} -e^{i (\beta_1+ \beta_3)} + e^{i(\beta_1+\beta_2+\beta_3)} - 1 \right) \int {dr \over r^3} A_{123},
%% \ee
Requiring the coefficient of this term be zero means $\beta_1+\beta_2+\beta_3 = 2 \pi n$.  

When this condition is satisfied, the correlator simplifies to
\begin{eqnarray*}
& \la V_{\beta_1}(z_1) V_{\beta_2}(z_2) V_{\beta_3}(z_3) \ra = \\
& \exp \left\{ -  \pi \gamma \left[ (1 - \cos \beta_1) (\ev_1 + \ev_{23}) + (1 - \cos
  \beta_2) (\ev_2 + \ev_{13}) +(1 - \cos \beta_3) (\ev_3 + \ev_{12}) \right] \right\} 
\end{eqnarray*}
The exclusive area integrals are given by, for example,
\bea
\ev_1 &=& \nev_1 - \nev_{12} - \nev_{13} + \nev_{123} \\
\ev_{23} &=& \nev_{23} - \nev_{123}
\eea
The formula for the triple overlap $\nev_{123}$ is somewhat
complicated, but it cancels in the 3-point function, because the
integrals appear in combinations such as
\be
\ev_1 + \ev_{23} = \nev_1 - \nev_{12} - \nev_{13} + \nev_{23}
\ee
Therefore the 3-point function simplifies to
\begin{eqnarray*}
& \la V_{\beta_1}(z_1) V_{\beta_2}(z_2) V_{\beta_3}(z_3) \ra = \\
& \exp \left\{ -  \pi \gamma \left[ (1 - \cos \beta_1) (\nev_1 -
    \nev_{12} - \nev_{13} + \nev_{23}) \right] \right\} \times {\rm
    (cyclic\  permutations)} 
\end{eqnarray*}
In terms of the weights $\Delta_i = {\pi \over 2} \gamma (1 - \cos \beta_i)$ the
3-point function can be written
\be
 \la V_{\beta_1}(z_1) V_{\beta_2}(z_2) V_{\beta_3}(z_3) \ra =
 \left| \exp \left\{- (\Delta_1  + \Delta_2 - \Delta_3)(I_1 - I_{12})
 \right\}\right|^2 \times {\rm
    (cyclic\  permutations)} 
\ee
The combination $(I_1 - I_{12})$  is exactly the same as in the 2-point
function, so
\be
\la V_{\beta_1}(z_1)  V_{\beta_2}(z_2) V_{\beta_3}(z_3) \ra =  \left| \left( {\delta \over z_1 - z_2} \right)^{\Delta_1 + \Delta_2 - \Delta_3}  \left( {\delta \over z_1 - z_3} \right)^{\Delta_1 + \Delta_3 - \Delta_2} 
 \left( {\delta \over z_2 - z_3} \right)^{\Delta_2 + \Delta_3 - \Delta_1} \right|^2
\ee

This is the 3-point function required by conformal invariance for three operators of weights $\Delta_i, \bar \Delta_i$, with $\Delta_i = \bar \Delta_i$.  It is worth noting that scale invariance alone is not enough to fix this form---scale invariance requires only that the total scaling dimension of any term on the right-hand side be consistent with the total scaling dimension of the fields in the correlator, but not this particular structure.  However global conformal invariance does require this form (in any number of dimensions), because there are no conformal invariants that can be built from less than 4 points. 

Starting from the Mobius invariance of the distribution
Eq. \ref{pdist} one could presumably prove that well-behaved
correlators must be of this form.  The statement is non-trivial
because of the issue of IR divergences;  correlators that depend on
the IR regulator will not in general respect this form.

\section{Four-point function}
In this section we find the 4-point function and analyze its
properties.
By now the procedure is familiar. The 4-point function is
\be
\begin{split}
 \la V_{\beta_1}(z_1)  V_{\beta_2}(z_2) V_{\beta_3}(z_3)
V_{\beta_4}(z_4) \ra = 
 \exp \{ - \pi \gamma [ \sum_i (1 - \cos \beta_i) \ev_i +
\sum_{i < j} (1 - \cos (\beta_i + \beta_j)) \ev_{ij} +  \\
 \sum_{i < j <
  k} (1 - \cos (\beta_i + \beta_j + \beta_k)) \ev_{ijk} + (1 -
\cos \sum_i \beta_i ) \ev_{1234} ] \}
\end{split}
\ee
The charge cancellation condition works as before: $\ev_{1234}$ is the
only infrared divergent quantity, so the correlator is zero unless
\be
\sum_i \beta_i = 2 \pi n
\ee
Using the charge cancellation condition, the 4-point function is
\be
\begin{split}
 \la V_{\beta_1}(z_1)  V_{\beta_2}(z_2) V_{\beta_3}(z_3)
V_{\beta_4}(z_4) \ra =  \\
\left| \exp \left\{ - \D_1 (\ev_1 + \ev_{234})
 - \D_2 (\ev_2 + \ev_{341}) - \D_3 (\ev_3 + \ev_{412}) - \D_4 (\ev_4 +
 \ev_{123}) 
- \sum_{i<j} \D_{ij} \ev_{ij} \right\} \right|^2
\end{split}
\ee
where we have defined
\be
\D_{ij} \equiv { \pi \over 2}\gamma (1 - \cos(\beta_i + \beta_j))
\ee
Note that due to the charge cancellation condition $\D_{12} =
\D_{34}$.

In the case of four points, the exclusive area integrals are given by
\bea \label{exc}
\ev_1 &=& \nev_1 - \nev_{12} - \nev_{13} - \nev_{14} + \nev_{123} +
\nev_{124} + \nev_{134} - \nev_{1234} \\
\ev_{12} &=& \nev_{12} - \nev_{123} - \nev_{124} + \nev_{1234}\\
\ev_{123} &=& \nev_{123} - \nev_{1234} \\
\ev_{1234} &=& \nev_{1234}
\eea
Using these relations and massaging the expression, the 4-point
function can be rewritten
\be
\begin{split}
& \la V_{\beta_1}(z_1)  V_{\beta_2}(z_2) V_{\beta_3}(z_3)
V_{\beta_4}(z_4) \ra = \\
& \left|\exp \left\{ -\sum_{i < j} (\D_i + \D_j - \D_{ij}) (\nev_1 -
\nev_{ij})- \left( \sum_i \D_i - {1 \over 2} \sum_{i<j} \D_{ij} \right)
\left( \sum_{i < j < k} \nev_{ijk} - 2 \nev_{1234} - 2 I_1 \right)
\right\} \right|^2
\end{split}
\ee
The first sum has the form of a product over six 2-point functions,
while the second term contains a nontrivial function of the positions
of the points. To be explicit,
\be
\begin{split}
& \la V_{\beta_1}(z_1)  V_{\beta_2}(z_2) V_{\beta_3}(z_3)
V_{\beta_4}(z_4) \ra = \\
&  \left| \prod_{i<j} \left(\delta \over z_{ij} \right)^{\D_i + \D_j -
  \D_{ij}} \exp \left\{- \left( \sum_i \D_i - {1 \over 2} \sum_{i<j} \D_{ij} \right)
\left( \sum_{i < j < k} \nev_{ijk} - 2 \nev_{1234} - 2 I_1 \right)
\right\} \right|^2
\end{split}
\ee
The interesting functional dependence on the location of the points is
all contained by the function $ f(z_1, z_2, z_3, z_4) = \sum_{i < j < k} \nev_{ijk} - 2
\nev_{1234} - 2 I_1$.

It is possible to compute the 4-point function in full generality and
show that it is conformally invariant. However, since the distribution
of bubbles is conformally invariant, the 4-point function is
guaranteed to be conformally invariant unless infrared divergences
arise. Therefore, we will assume conformal invariance and compute the
4-point function with the four points at 
\bea
z_1 &=& z \\
z_2 &=& 0 \\
z_3 &=& 1 \\
z_4 &=& \infty
\eea
With this assumption, the four-point function is an infinite constant
times a nontrivial function of $z$,
\be
\la V_{\beta_1}(z)  V_{\beta_2}(0) V_{\beta_3}(1)
V_{\beta_4}(\infty) \ra =
C \left|  z^{-(\D_1 + \D_2 - \D_{12}) }(1 - z)^{-(\D_1 + \D_3 - \D_{13})}
\exp  \left\{- \left( \sum_i \D_i - {1 \over 2} \sum_{i<j} \D_{ij} \right)
 \nev_{123} 
\right\} \right|^2
\ee
It now remains to evaluate the integral $\nev_{123}$.

\subsection{Evaluation of the triple overlap integral}
We need to evaluate
\be
\nev_{123} \equiv {1 \over \pi} \int {dr \over r^3} A_{123}
\ee
where $A_{123}$ is the area contained within the triple overlap of
three disks of radius $r$ centered at the points $z_1, z_2$, and $z_3$.

To evaluate this we will need a formula for the area of triple overlap of three circles.  In some cases this reduces to a double overlap, but in situations where the triple overlap is of triangular type (e.g. a region bounded by the arcs of three distinct circles) the area is \cite{kratky}:
\be
A_{123} = {1 \over 2} \left(A_{12} + A_{13} +A_{23}  - \pi r^2 \right) + A_T,
\ee
where $A_T$ is the area of the triangle with vertices at the three
points. We will do the computation for triangles for which the triple
overlap area is always given by this formula for any disk size $r$,
which amounts to assuming that the triangle is sufficiently close to
equilateral, with none of the angles exceeding $90^o.$ However, our
final formula will be valid for any arrangement of three points.

\begin{figure}
\begin{center}
\includegraphics[scale=.3, angle=-90]{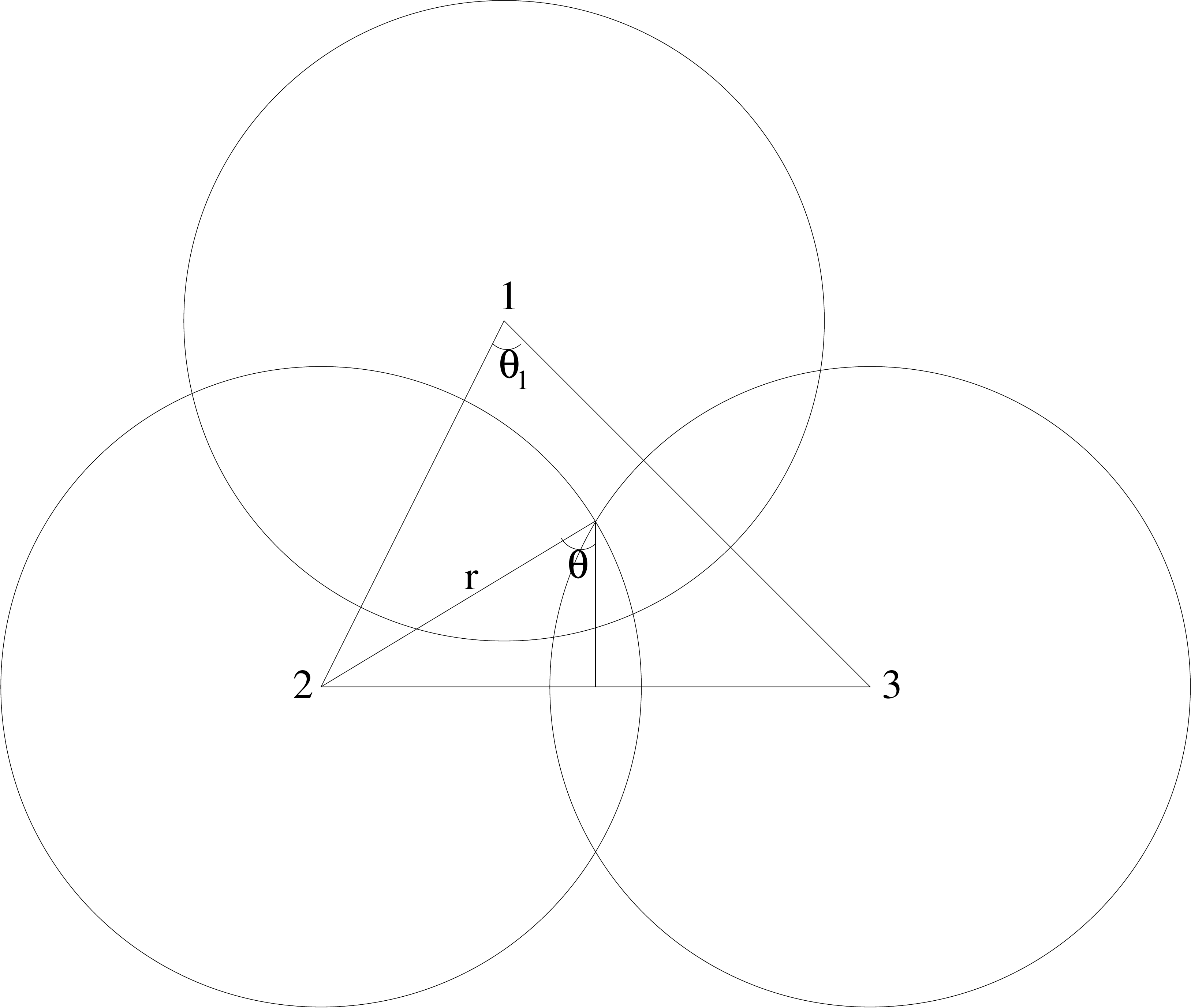}
\end{center}
\caption{It is convenient to change variables in the integration,
  using $\theta$ shown in the figure instead of $r$ as the integration variable.}
\label{triple2}
\end{figure}

\begin{figure}
\begin{center}
\includegraphics[scale=.3, angle=-90]{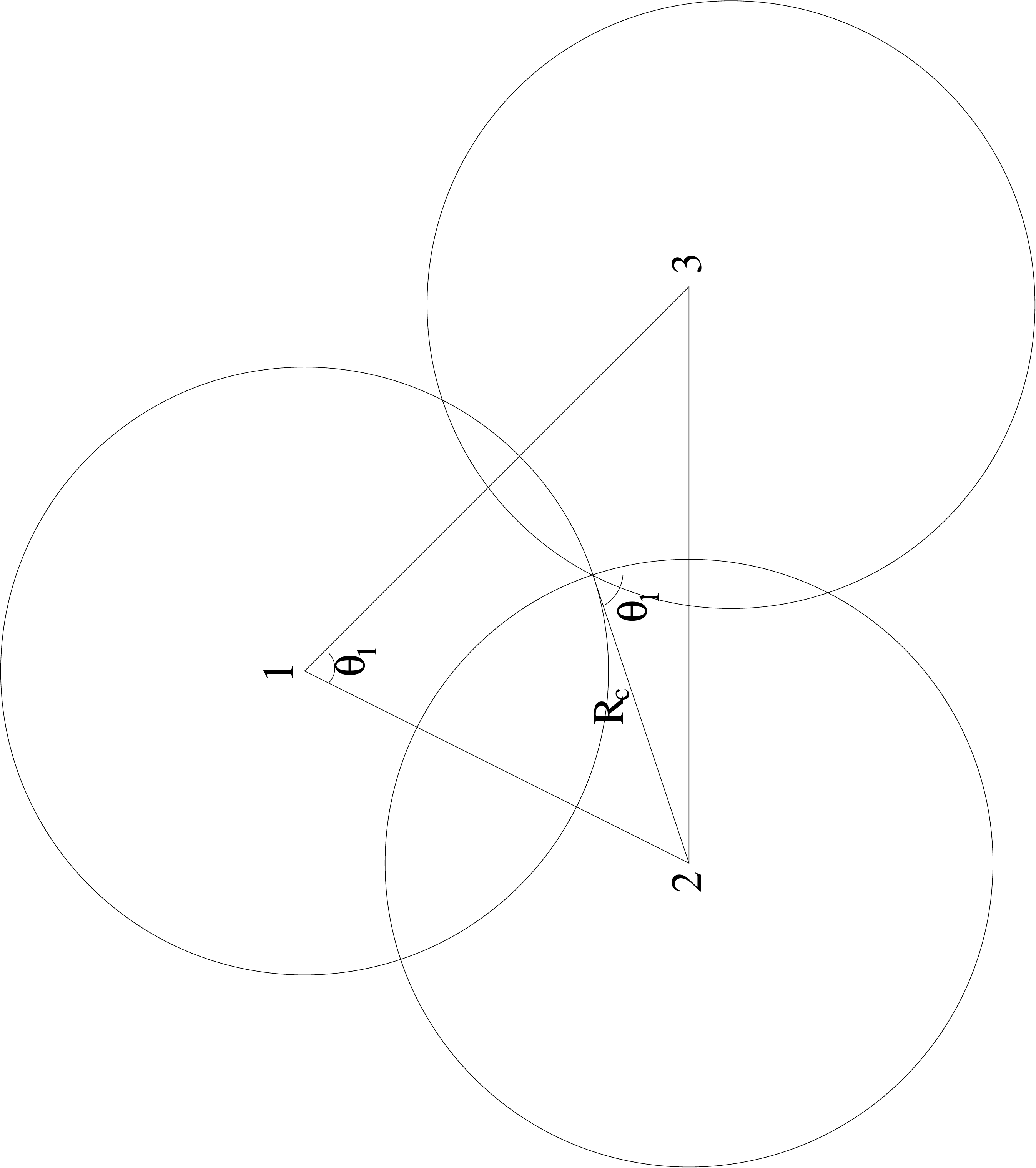}
\end{center}
\caption{The smallest radius disk which can cover all three points is
  defined by $r=R_c$. The figure shows that the lower limit if
  integration corresponds to $\theta = \theta_1$.}
\label{triple1}
\end{figure}

The triple overlap integral begins to be nonzero at the smallest value
of $r$ such that a disk can cover all three points. This value is
called the circumradius $R_c$. 
By subtracting the area inside a wedge of the circle from a triagular
area (see figure \ref{triple2}), we find
\be
{1 \over 2} A_{23} = r^2 (\pi/2 - \theta) - r^2 \sin \theta \cos \theta  
\ee  
where $\theta$ is the angle shown in the figure.
The integral of $A_{23}$ is infrared divergent, but the quantity
$A_{23} - \pi r^2$ is infrared safe, 
\be
{1 \over 2} (A_{23} - \pi r^2) =   -(\theta + \sin \theta \cos \theta)
r^2
\ee
so we focus on it.
We need to integrate the triple overlap from the lower limit $R_c$
where it is first nonzero.  So the
$A_{23}$ piece of the integral is given by
\be
{1 \over 2} \int_{R_c}^R {dr \over r^3} ( A_{23} - \pi r^2) = 
- \int_{R_c}^R {dr \over r} \left[  \theta + \sin \theta \cos
\theta \right]
\ee
and we can now freely take $R \to \infty$ because the integral is
finite in the infrared.
The angle $\theta$ is related to $r$ by 
\be
\sin \theta = {d_{23} \over 2 r}
\ee
Also, as shown in figure \ref{triple1}, the lower limit $r = R_c$
corresponds to an upper limit on $\theta$, $\theta = \theta_1$, where
$\theta_1$ is the angle of the triangle with its vertex at point 1. 
The upper limit $r = \infty$ corresponds to $\theta = 0$.

Making the change of variables, the integral is
\be
{1 \over 2} \int_{R_c}^\infty {dr \over r^3} (A_{23} - \pi r^2) = 
- \int_0^{\theta_1} d\theta \left[ \theta \cot \theta +
  \cos^2 \theta \right]
\ee
We can now write the full integral.
\be
\int_{R_c}^{R}{dr \over r^3} A_{123} =  \int_{R_c}^R {dr \over r^3} \left[ {1 \over 2} \left( (A_{12} -
\pi r^2) + (A_{13} - \pi r^2) + (A_{23}- \pi r^2) \right) + \pi r^2 +
A_T \right]
\ee
Now the only infrared divergent term is the trivial $\pi r^2$; for the
rest of the terms we take $R \to \infty$ to get
\be
\int_{R_c}^{R} {dr \over r^3} A_{123} = \pi \log{R \over R_c} + {A_T \over 2 R_c^2}
-\sum_i \int_0^{\theta_i} (\theta \cot \theta + \cos^2 \theta) d\theta
\ee
where the sum is over the three angles of the triangle.

The integral gives
\be
 \int_0^{\theta_i} (\theta \cot \theta + \cos^2 \theta) d\theta = 
\theta_i \ln(\sin \theta_i) + \theta_i ({1 \over 2} + \ln 2) + {1 \over 4}
\sin(2 \theta_i) + {1 \over 2} \Im \left[\li (e^{2 i \theta_i})\right]
\ee
where $\li$ is the dilogarithm function,
\be
\li (z) \equiv \sum_{n=1}^\infty {z^n \over n^2}
\ee
We would like to write everything in terms of the angles and lengths
of the sides. By examining figure \ref{triple1}, we find
\bea
2 R_c &=& {d_{23} \over \sin{\theta_1}} = {d_{13} \over \sin{\theta_2}} =
{d_{12} \over \sin{\theta_3}} \\
A_T &=& R_c^2 \sum_i \sin \theta_i \cos \theta_i 
\eea
We can now rewrite the integral as
\be
\begin{split}
\pi I_{123} =  {\pi \over 3} \ln \left( {8R^3 \sin \theta_1 \sin \theta_2
  \sin \theta_3 \over d_{12} d_{13} d_{23}} \right) + {1 \over 2}
  \sum_i \sin \theta_i \cos \theta_i
 \\ 
- 
\sum_i \left( \theta_i \ln(\sin \theta_i) + \theta_i ({1 \over 2} + \ln 2) + {1 \over 4}
\sin(2 \theta_i) + {1 \over 2} \Im \left[\li (e^{2 i \theta_i})\right]
\right)
\end{split}
\ee
In combining the terms, some simplifications occur because $\sum
\theta_i = \pi$. A nice symmetric way to write the answer is
\be
\pi I_{123} = -{\pi \over 2} + {\pi \over 3} \ln {R^3 \over \left| z_{12} z_{13}
  z_{23} \right| } -
\sum_i \left( (\theta_i - {\pi \over 3}) \ln (\sin \theta_i) + {1 \over
  2} \Im \left[ \li(e^{2 i \theta_i})\right] \right)
\ee
The sum is over all three angles of the triangle formed by the
three points, and each angle is defined in the conventional way so
that $0 \leq \theta_i \leq \pi$. Although our derivation is only valid
for triangles which are sufficiently close to equilateral, the answer
written in this way is valid for any arrangement of the three points.

The answer can also be written in the pleasing form
\be
\pi I_{123} = -{\pi \over 2} + \pi \ln R - \theta_1 \ln d_{23} -
\theta_2 \ln d_{31} - \theta_3 \ln d_{12}
- {1 \over 2} \sum_i \Im \left[ \li(e^{2 i \theta_i})\right] 
\ee
It is particularly simple when the three points are on a line. If
point 2 is between points 1 and 3, then $\theta_1 = \theta_3 = 0$ and
$\theta_2 = \pi$, so that
\be
I_{123} = -1/2 + \ln R - \ln d_{31} \ \ \ \ {\rm (collinear)}
\label{collinear}
\ee
because all of the dilogarithms vanish.

Having written the answer in terms of the angles, we want to write it
as a function of $z$ for our special choice of points,
\bea
z_1 = z \nonumber \\
z_2 = 0 \nonumber \\
z_3 = 1 \nonumber \\
\eea
The angles should satisfy $0 \leq \theta_i \leq \pi$. For $\Im(z) > 0$
we have
\bea
e^{2 i \theta_1} &=& {\bar z (1 - z) \over z (1 - \bar z)} \\
e^{2 i \theta_2} &=& {z \over \bar z} \\
e^{2 i \theta_3} &=& {1 - \bar z \over 1 - z}
\label{thetaz}
\eea
The above is not valid for $\Im(z) < 0$, but it is clear that the
correct answer should be invariant under $z \to \bar z$, so we will
just work it out for $z$ in the upper half plane and determine the
value in the lower half plane by symmetry.

We rewrite the answer piece by piece in terms of $z$. 
To write the dilogarithm part of the answer in terms of $z$, it is
convenient to rewrite the answer in terms of the Bloch-Wigner function
$\dtwo$. The relation is
\be
\dtwo(z) = \Im[\li(z)] + \arg(1 - z) \ln |z|
\label{lid}
\ee
so that $\dtwo(e^{i \theta}) = \Im[\li(e^{i \theta})]$.
The combination which appears in the answer is
\be
{1 \over 2} \sum_i \dtwo \left(e^{2 i \theta_i} \right)
 = {1 \over 2} \left\{ \dtwo \left( z \over \bar z \right) + \dtwo\left( \bar z
  (1 - z) \over z (1 - \bar z) \right) + \dtwo \left(1 - \bar z \over
  1 - z \right)\right\} = \dtwo(z)
\ee
where we have used an identity of the Bloch-Wigner function to get the
last equality (see \cite{lewin} section 7.2).

For the part not
involving dilogarithms, we need to solve for $\theta_i$ in terms of
$z$. Again for $z$ in the upper half plane, we can invert
(\ref{thetaz}) to get
\bea
i \theta_2 &=& \ln \left( z \over |z| \right) \nonumber \\
-i \theta_3 &=& \ln \left(1 - z \over |1 - z| \right) 
\eea
subject to the usual convention that the branch cut for the logarithm
is taken to be on the negative real axis.
 $\theta_1$ is not needed because in the answer it multiplies $\ln
d_{23}$, which is zero.
Plugging these in and simplifying, 
\be
 - \theta_1 \ln d_{23} -
\theta_2 \ln d_{31} - \theta_3 \ln d_{12} = 
 - \Im\left[ \ln z \ln ( 1 - \bar z)
 \right]
\ee

So the answer as a function of $z$ is
\be
\pi \nev_{123} = - {\pi \over 2} + \pi \ln R - \Im[\ln z \ln(1 - \bar
z)] - \dtwo(z) \ \ \ {\rm for } \ \Im(z) > 0
\ee
% The dilogarithm part of our answer is
% \be
% \begin{split}
% \Im \left[ \li \left(e^{2 i \theta_1} \right) +\li \left(e^{2 i
%     \theta_2} \right) +\li \left(e^{2 i \theta_3} \right) \right] = 
% \Im \left[ \li \left(e^{2 i (\pi - \theta_2 - \theta_3)} \right) +\li \left(e^{2 i
%     \theta_2} \right) +\li \left(e^{2 i \theta_3} \right) \right] = \\
% \Im \left[ - \li \left(e^{2 i ( \theta_2 + \theta_3)} \right) +\li \left(e^{2 i
%     \theta_2} \right) +\li \left(e^{2 i \theta_3} \right) \right]
% \end{split}
% \ee
% This can be simplified by using the Abel identity for the dilogarithm,
% \be
% \li(u) + \li(v) - \li(uv) = \li \left( u - uv \over 1 - uv \right) +
%  \li \left( v - uv \over 1 - uv \right) + \ln \left( 1 - u \over 1 -
%  uv \right) \ln \left( 1 - v \over 1 -
%  uv \right)
% \ee
% Using the Abel identity and the relation between $z$ and the
% $\theta_i$ this becomes
% \be
% \Im \left[ \li \left(e^{2 i \theta_1} \right) +\li \left(e^{2 i
%     \theta_2} \right) +\li  \left(e^{2 i \theta_3} \right) \right] = 
% \Im \left[ \li(z) + \li(1 - \bar z) + \ln \bar z \ln (1 - z) \right]
% \ee

Now our algebra has been done under the assumption that
$z$ is in the upper half plane, but the answer must be symmetric under
$z \to \bar z$. The answer as it stands is a constant term plus two
functions which are odd under $z \to \bar z$. Therefore, the correct
answer has an additional factor of the sign of the imaginary part of z,
\be
 \pi  \nev_{123} = - {\pi \over 2} + \pi \ln R  - {\rm sgn}(\Im(z)) \left\{ \Im[ \ln z
  \ln(1 - \bar z)]  +  \dtwo(z)  \right\}
\ee

\subsection{Properties of the 4-point function}
Plugging this in, we have the 4-point function 
\bea \label{4pt3}
\la V_{\beta_1}(z)  V_{\beta_2}(0) V_{\beta_3}(1)
V_{\beta_4}(\infty) \ra =
C \left| z^{-\D_1 - \D_2 + \D_{12}} (1 - z)^{-\D_1 - \D_3 + \D_{13}}
 \right|^2 \times \nonumber \\
 \exp  \left\{  \left( {2 \over \pi} \sum_i \D_i - {1 \over \pi} \sum_{i<j} \D_{ij} \right)
 {\rm sgn}(\Im(z)) \left\{ \Im[\ln z \ln (1 - \bar z)] + \dtwo(z) \right\}
\right\}
\eea
 By conformal invariance, aside from simple prefactors which depend on
 the conformal weights of the fields, the 4-point function can depend
 only on the cross ratio. For our choice of points the cross ratio is
 just $z$: 
\be
z = {(z_1 - z_2) (z_4 - z_3) \over (z_3 - z_2) (z_4 - z_1)}~.
\ee

The 4-point function in a conformal field theory should be crossing-symmetric---it should be invariant under interchanging the points where the operators are inserted.  
Interchanging the points corresponds to the following
group of transformations on $z$:
\be
z \to 1 - {1 \over z} \to {1 \over 1-z} \to {1 \over z} \to 1 - z \to
{-z \over 1 - z}
\ee
The last three are odd permutations of the four points;
the rest are even
\cite{zagier}. The Bloch-Wigner function $\dtwo(z)$ has the property that
it changes sign under odd permutations and is invariant under
even permutations  \cite{zagier}. Since the sgn function changes sign
only under odd permutations,  ${\rm
  sgn}(\Im(z)) \dtwo(z)$ is invariant under all permutations.

The other nontrivial factor in
the 4-point function, (${\rm sgn}(\Im(z)) 
\Im [ \ln z \ln (1-z)]$), is slightly more
complicated, but one can check that under permutations it transforms
in the correct way to contribute the right factors. To
take a nontrivial example, under $z \to 1/z$ this function transforms
as
\be
{\rm sgn}(\Im(z^{-1})) 
\Im [ \ln z^{-1} \ln (1-\bar z^{-1})] = {\rm sgn}(\Im(z)) 
\Im [ \ln z \ln (1-\bar z)] - \pi \ln |z|.
\ee
One can use this to check that the 4-point function Eq. \ref{4pt3} satisfies
\be
\la V_{\beta_1}(z)  V_{\beta_2}(0) V_{\beta_3}(1)
V_{\beta_4}(\infty) \ra = z^{-2 \Delta_1} \bar z^{-2 \Delta_1}
\la V_{\beta_1}(1/z)  V_{\beta_4}(0) V_{\beta_3}(1)
V_{\beta_2}(\infty) \ra,
\ee
which is the correct behavior \cite{cft}.

The four-point function simplifies when all 4 points are on a line, or
more generally a circle. Using the simple form (\ref{collinear}) for
the triple overlap integral, for $z$ real and negative we have
\bea
\la V_{\beta_1}(z)  V_{\beta_2}(0) V_{\beta_3}(1)
V_{\beta_4}(\infty) \ra =  C \left| z^{-\D_1 - \D_2 + \D_{12}} (1 - z)^{-\D_1 - \D_3 + \D_{13}}
 \right|^2 \times \nonumber \\
 \exp  \left\{  \left( 2 \sum_i \D_i - \sum_{i<j} \D_{ij} \right) \ln|1-z|
\right\} \ \ \ \ (z \ {\rm real\ and\ negative})~.
\eea
Therefore for all 4 points on a circle, the 4-point function is simply
a product of power laws; the powers which appear depend on the order
of the points.

However, there is a problem.  Without the sgn
factor, the functions in the exponent in Eq. \ref{4pt3} would be 
odd under $z \to \bar z$. Furthermore, they are real analytic functions away from lines of discontinuity running along the real axis from $-\infty$ to $0$ and from $1$ to $+\infty$.  So before multiplying by the sign function, the exponent is a real analytic function in a finite region around $z = 1/2$. Therefore after including the  sign the 4-point function is not real analytic  as we
drag $z$ across the real axis near $z = 1/2$, even though $z$ is separated from the other
points at 0 and 1 by a finite distance. In a conventional field theory, correlation functions should be analytic except when two points approach each other. 

The nonanalyticity of the 4-point function indicates that the system
we have defined does not correspond to a full-fledged conformal field
theory. This is not too surprising, because we have made a number of
approximations in treating the physics of bubble nucleation; one may still
expect that our system can be obtained as the limit of a genuine conformal field theory. 
We will
discuss this further in the conclusions. 
% For the purposes of expanding at small $z$, it is convenient to
% convert back to the polylogarithm, since it has a simple Taylor
% expansion. The relation (\ref{lid}) gives
% \be
% \dtwo(z) = \Im \left[ \li(z) \right] + {1 \over i} \ln \left( 1 - z
%   \over |1 - z| \right) \ln |z| =  \Im \left[ \li(z)  + {1 \over
%   2} \ln (1 - z) \ln |z|^2 \right]
% \ee
% As a result, the 4-point function can be written
% \be
% \pi \nev_{123} = - {\pi \over 2} + \pi \ln R  - {1 \over 2} \left| \Im \left[ 2 Li(z) + \log z
%    \log|1 - z|^2 \right] \right|
% \ee

\section{Central charge}

Conformal field theories on curved spaces have a conformal anomaly.  Specifically, in 2D CFTs $T^a_{~a} = -{c \over 12} {\hat R}$, where $\hat R$ is the scalar curvature of the 2D space.   Since the trace of the stress tensor is related to the variation of the action with respect to the conformal factor in the metric, one can compute $c$ 
by taking the derivative of $\ln Z$ with respect to the log of the curvature.  On a sphere of radius $R$, 
\be
Z(R) = R^{c/3} Z_0,
\ee
where $Z_0$ is the partition function on a sphere of unit radius \cite{zam}.

Given the results of the previous section, it is at best unlikely that the theory as we have defined it is a full conformal field theory.  Nevertheless we will proceed, as we can easily compute $Z$ on a sphere. The partition function  is
\be
Z = \sum_{n=1}^\infty {\gamma^n \over n!} \int d\Omega_2 \left(  \int_\epsilon^{\pi - \epsilon} {d \psi \over \sin^3 \psi} - \Lambda \right) ,
\ee
where $\epsilon$ is a cutoff on disk angular size and $\Lambda$ is a (cosmological) constant added as a local counterterm to cancel the leading UV divergence from small $\epsilon$ (as a constant multiplicative factor in $Z$ it cancels out of all the correlators computed earlier).  Computing the integral and restoring the dimensions gives
\be
\ln Z = 4 \pi \gamma \left( {R^2 \over \epsilon^{2}} + \ln {R \over \epsilon} + \ln 2 - {1 \over 6} - \Lambda R^2 + {\cal O}(\epsilon^2)\right).
\ee
Setting the counterterm $\Lambda = 1/\epsilon^2$ cancels the quadratic UV divergence, but the log is an anomaly that cannot be cancelled with any local counterterm.  This is precisely what one expects for a 2D CFT on a sphere with central charge
\be
c = 12 \pi \gamma ~.
\ee

For $c<1$, unitarity implies that $c$ must take a discrete set of values (the minimal models).  Therefore if this calculation is taken seriously it indicates that our model cannot be unitary at small $\gamma$.  However it is worth mentioning that we have included neither perturbative fluctuations of the field nor of the geometry.  One expects graviton fluctuations to contribute a term of order $(M_P/H)^{d-2}$, and so it is possible that the term we have computed is only one of several contributions to the central charge of a putative complete theory.  At large $\gamma$, it is reasonable to expect that the instanton fluctuations are the dominant contribution, and indeed in that limit the theory becomes free.  It would be interesting to compare this limit to the analogue in AdS space, where one would take $G_N \to 0$ with the AdS radius held fixed so that $c \to \infty$.

\subsection{CFT on a fractal}

In the string theory landscape there are so-called ``terminal vacua": minima with either zero or negative cosmological constant.  A region which tunnels to one of these minima has at most a finite probability of nucleating any more bubbles before infinite time (for a zero CC bubble) or  a big crunch (for a negative CC bubble).  How best to deal with these regions is unclear, but the proposal of 
 \cite{gv} is to excise them and attempt to define a CFT on the remaining space, perhaps including lower dimensional defect CFTs on the boundaries.
 
A simple toy model for terminal vacua in our 2D CFT is to assume there is some rate $\gamma$ to produce ``dead" disks.  Following the suggestion of \cite{gv} then corresponds to computing correlators in the regions outside these dead disks.  Since the region covered by zero disks of any type is a fractal set of measure zero, these correlators will ``live" on a fractal set.

One immediate problem is that when $\gamma$ is large this set is not just measure zero, but empty (recall that $d_F = d-\gamma \Omega_d$, and when $d_F \leq 0$ the set is empty).  One can consider the case of $\gamma < 1$ and continue, but it is clear that the resulting theory will have a very different structure than what we have considered so far. 

Since the set is by definition $N(z)=0$ one cannot compute correlators
of the disk number operator in the way we have been proceeding.  There
are two obvious approaches one could take to this.  One is to compute
the probability that some set of points $z_i$ in the full space are
all in the set, or various conditional probabilities (such as the odds
that if one point is in the set, the others are as well).   Computing
these probabilities is not difficult using the techniques we have
already developed, and the results depend on the distances between the
points.  However proceeding in this manner we have not succeeded in
defining a set of probabilities that are independent of the IR cutoff.
The problem in a nutshell is that the number of disks which covers
some but not all of a certain set of points is IR finite when
integrated against the distribution, but on the other hand the number
of disks that would cover all of the points is infrared divergent.  

Another approach is to consider more types of disks:  a ``dead" type which defines the fractal, and then one or more other ``live" types.  One could then try to compute correlators of the number operator for ``live" disks within the set of points covered by zero ``dead" disks.  However this analysis requires taking into account interactions between the different types of bubbles, something which we will not consider in this note.

%   For example, one could regard regions with $|N|>n$ as terminal vacua which have been excised, and then compute correlators on the remaining set (which will be a fractal of measure zero).

%Such correlators can be defined using the same techniques described in the rest of the paper.  One could, for example, ask for the conditional probability that if $z$  and $z'$ are in the set $|N|>n$, what are the odds that $N(z)=N(z')$?

%FINISH THIS, COMMENT HOW THESE CORRELATORS ARE IR DIVERGENT

\section{Generalization to arbitrary dimension}
The definition of the theory can be easily generalized to arbitrary
dimensions. The partition function in $d$ dimensions is
\be
Z = Z_+ Z_-
\ee
with
\be
Z_+ = \exp \left( \gamma_+ \int_\delta^R {dr \over r^{d+1}} \int d^d x
\right)
\ee
We will show that the two and three point functions
generalize in a simple way. To make the discussion here easier to
follow, we include some formulas and discussion which overlap with the
earlier sections of the paper.

The general N-point function is
\begin{eqnarray}
\la e^{i \beta_1 N(z_1)} e^{i \beta_2 N(z_2)} ... e^{i \beta_n N(z_n)} \ra =
\exp (-2 \dvol \gamma [ \sum_{i = 1}^n (1 - \cos \beta_i) \ev_i + \sum_{i
    < j}(1 - \cos (\beta_i + \beta_j) ) \ev_{ij} + \nonumber \\
 \sum_{i < j < k} 
(1 - \cos (\beta_i + \beta_j + \beta_k) ) \ev_{ijk} + ... ])
\end{eqnarray}
where now $z_i$ is a point in $d$-dimensional space and $\dvol$ is the
volume of a unit $d$-sphere.
As before, define the integral
\be
\ev_1 \equiv {1 \over \dvol} \int {dr \over r^{d+1}} \ea_1 (r, z_i)~.
\ee

The two-point function when the charge conservation condition is
satisfied is 
\be
\la e^{i \beta N(z_1)} e^{-i \beta N(z_2)} \ra = 
\exp (- \dvol \gamma (1 - \cos \beta) (\ev_1 + \ev_2))
\ee
Now $\ev_1 = \ev_2 = \nev_1 - \nev_{12}$ as before. 
The 2-point function is then
\be
\la e^{i \beta N_1} e^{-i \beta N_2} \ra = 
\exp (-2 \dvol \gamma (1 - \cos \beta) (\nev_1 - \nev_{12}))
\ee
The integral is given by
\be
\nev_1 - \nev_{12} = {1 \over \dvol} \int_\delta^R {dr \over r^{d+1}} A_1(r) -  
{1 \over \dvol} \int_{d_{12}/2}^R {dr \over r^{d+1}} A_{12}(r, d_{12}) 
\ee
It is helpful to rewrite this as
\be
\nev_1 - \nev_{12} = {1 \over \dvol} \int_\delta^{d_{12} \over 2} 
{dr \over r^{d+1}} A_1(r) -  
{1 \over \dvol} \int_{d_{12} \over 2}^R {dr \over r^{d+1}} \left[ A_1(r) -A_{12}(r, d_{12}) \right]
\ee
The first term can be integrated immediately using $A_1(r) = \dvol
r^d$ to get
\be
\nev_1 - \nev_{12} = \ln \left( d_{12} \over 2 \delta \right) -  
{1 \over \dvol} \int_{d_{12} \over 2}^R {dr \over r^{d+1}} \left[ A_1(r) -A_{12}(r, d_{12}) \right]
\ee
The remaining integral can be evaluated explicitly, but the crucial
information can be extracted more cheaply. The first term, the
integral of $A_1$, diverges logarithmically at large $r$. However, the
second term cancels this divergence, because the double overlap region
$A_{12}$ asymptotically has the same area as a single disk,
\be
A_{12} \to \dvol r^d \ \ {\rm as}\ r \to \infty~.
\ee
Therefore, the integral is infrared finite, and we can take $R \to
\infty$, so that the integral is
\be
{1 \over \dvol} \int_{d_{12} \over 2}^\infty {dr \over r^{d+1}} \left[ A_1(r) -A_{12}(r, d_{12}) \right]
\ee
This formula depends only on the dimensionful quantity $d_{12}$, but
dimensional analysis shows that the answer must be
dimensionless. Therefore it is a constant independent of the distance
$d_{12}$. So finally 
\be
\nev_1 - \nev_{12} = \ln \left( d_{12} \over \delta \right) - \nd
\ee
where $\nd$ is a dimension-dependent constant.

Plugging this in, we have
\be
\la e^{i \beta N(z_1)} e^{-i \beta N(z_2)} \ra = e^{ - 2
\dvol \gamma (1 -
  \cos \beta) \nd } \left(
  \delta \over |z_{12}| \right)^{2 \dvol \gamma (1 - \cos \beta)}
\ee
The prefactor can be absorbed into a redefinition of the ultraviolet
cutoff $\delta$. 
The dimension of the vertex operator $e^{i \beta N}$ is
\be
\Delta = \bar{\Delta} =  {\dvol \over 2} \gamma (1 - \cos \beta)
\ee

Having found the dimensions we can rewrite the general n-point
function,
\be \label{general}
\la e^{i \beta_1 N(z_1)} e^{i \beta_2 N(z_2)} ... e^{i \beta_n N(z_n)} \ra =
\left| \exp \left\{ - \sum_{i = 1}^n \Delta_i \ev_i - \sum_{i
    < j} \Delta_{ij} \ev_{ij} -
 \sum_{i < j < k} 
\Delta_{ijk} \ev_{ijk} - ... \right\} \right|^2
\ee

Now to evaluate the 3-point function explicitly. In general, we have
\be
\la e^{i \beta_1 N(z_1)} e^{i \beta_2 N(z_2)} e^{i \beta_3 N(z_3)} \ra =
\left| \exp \left\{ -  \Delta_1
\ev_1 - \Delta_2 \ev_2 - \Delta_3 \ev_3 - \Delta_{12} \ev_{12} - \Delta_{13}
\ev_{13} - \Delta_{23} \ev_{23} - \Delta_{123} \ev_{123}] \right\} \right|^2
\ee
The correlation function is zero unless $\Delta_{123} = 0$ because
$\ev_{123}$ is infrared divergent. $\Delta_{123} = 0$ when the charge
conservation condition is satisfied, $\beta_1 + \beta_2 + \beta_3 =
2 \pi n$ with $n$ an integer. Using the charge conservation condition,
we have relations like $\Delta_{12} = \Delta_3$. Also, recall that the
exclusive volumes are given by
\begin{eqnarray}
\ev_1 &=& \nev_1 - \nev_{12} - \nev_{13} - \nev_{23} + \nev_{123} \\
\ev_{12} &=& \nev_{12} - \nev_{123}
\end{eqnarray}
so the 3-point function becomes
\be
  \left| \exp \left\{ - \Delta_1
(\nev_1 - \nev_{12} - \nev_{13} + \nev_{23}) - \Delta_2
(\nev_2 - \nev_{12} - \nev_{23} + \nev_{13}) - \Delta_3
(\nev_3 - \nev_{32} - \nev_{13} + \nev_{12})]) \right\} \right|^2
\ee
Note that the triple overlap region $\nev_{123}$ does not
appear. Collecting terms, we have
\be
\left| \exp \left\{-(\Delta_1 +
\Delta_2 - \Delta_3)(\nev_1 - \nev_{12}) - (\Delta_1 +
\Delta_3 - \Delta_2)(\nev_1 - \nev_{13}) - (\Delta_2 +
\Delta_3 - \Delta_1)(\nev_2 - \nev_{23})\right\} \right|^2
\ee
where we have used $\nev_1 = \nev_2 = \nev_3$. But this factorizes
into 2-point functions! This is precisely the form the 3-point
function must take due to conformal invariance.
Explicitly, it is
\be
\la e^{i \beta_1 N(z_1)} e^{i \beta_2 N(z_2)} e^{i \beta_3 N(z_3)} \ra =
({\delta^2 \over z_{12} \bar
z_{12}})^{\Delta_1 + \Delta_2 - \Delta_3}({\delta^2 \over z_{13} \bar
z_{13}})^{\Delta_1 + \Delta_3 - \Delta_2})({\delta^2 \over z_{23} \bar
z_{23}})^{\Delta_2 + \Delta_3 - \Delta_1}
\ee
where we have absorbed the same constant factor into the definition
$\delta$ as in the 2-point function.

Now for the 4-point function. Performing a similar analysis as for the
3-point function, we find
\begin{eqnarray}
\la e^{i \beta_1 N_1} e^{i \beta_2 N_2} e^{i \beta_3 N_3} e^{i \beta_4
  N_4} \ra =
\left| \exp \left\{ - \sum_{i < j} (\Delta_i + \Delta_j -
\Delta_{ij})(\nev_1 - \nev_{ij})\right\} \right|^2 \times \nonumber \\
\left| \exp  \left\{ -C_\Delta(\nev_{123} + \nev_{124}
+ \nev_{134} + \nev_{234} - 2 \nev_{1234} - 2 \nev_1) \right\} \right|^2
\end{eqnarray}
with the definition
\be
C_\Delta =  \Delta_1 + \Delta_2 + \Delta_3 +
\Delta_4 - \Delta_{12} - \Delta_{13} - \Delta_{14} \ .
\ee
The first factor has the form of factorized two point functions, so we
can rewrite this as
\begin{eqnarray}
\la e^{i \beta_1 N_1} e^{i \beta_2 N_2} e^{i \beta_3 N_3} e^{i
  \beta_4 N_4} \ra =
{\displaystyle \prod_{i < j}} \left(\delta^2 \over |z_{ij}|^2 \right)^{\Delta_i +
  \Delta_j - \Delta_{ij}} \times \\
\left| \exp \left\{- C_\Delta (\nev_{123} + \nev_{124}
+ \nev_{134} + \nev_{234} - 2 \nev_{1234} - 2 \nev_1) \right\} \right|^2 
\end{eqnarray}
So the first part of the 4-point function consists of simple power
laws.  The last line has all of the
interesting information in it, and involves the triple and quadruple
overlaps. This term is IR finite on its own, but it does depend on the
UV cutoff through $\nev_1$; this dependence is trivial.

Computing the 4-point function explicitly is a nontrivial task which we have only
accomplished in $d=2$ so far.

\subsection{Free field limit}

Our correlators have a free-field limit when the tunneling rate
becomes large. In taking $\gamma \to \infty$, the expectation value of
$N$ will become very large. Operators which are well-defined in this
limit should have $\beta_i \to 0$.  More preciesly, defining
\be
 \beta = \alpha \sqrt{2 \over \pi \gamma}
\ee
we want to take the limit $\gamma \to \infty$ with $\alpha$ fixed.

  To see this that the correlators factorize in this limit, start from the formula for an n-point correlator in general dimension, Eq. \ref{general}:
\be \label{general2}
\la e^{i \beta_1 N(z_1)} e^{i \beta_2 N(z_2)} ... e^{i \beta_n N(z_n)} \ra =\left| \exp \left\{ - \sum_{i = 1}^n \Delta_i \ev_i - \sum_{i
    < j} \Delta_{ij} \ev_{ij} -
 \sum_{i < j < k} 
\Delta_{ijk} \ev_{ijk} - ... \right\} \right|^2.
\ee
The exclusive integrated areas $I_{ijk...}^0$ satisfy the obvious generalizations of Eqs. \ref{exc}:
\be
I_{i_1 i_2 ... i_k}^0 = \sum_{l=0}^{n-k} {(-1)^l \over l!} \sum_{j_1,
  j_2, ..., j_l} I_{i_1 i_2 ... i_k j_1 ... j_l}
\ee
where the $j$ indices are summed from $1$ to $n$ and the factor of $l!$ corrects for overcounting.  In this expression we have defined $I_{ijk...}=0$ if any $i,j,k...$ are equal.
Then the exponent in Eq. \ref{general2} can be written
\be
\sum_{k=1}^n {\Delta_{i_1...i_k} \over k!} \sum_{l=0}^{n} {(-1)^l
  \over l!} \sum_{j_1,
  j_2, ..., j_l} I_{i_1 i_2 ... i_k j_1 ... j_l}
\ee
where we have also defined all areas with $m$ indices $I_{i_1i_2...i_m}=0$ if $m>n$.

To continue, we would like to evaluate the coefficient of the areas $I_{i_1i_2...i_l j_1...j_{m-l}}$ with $m$ indices, which is:
\be \label{termx}
\sum_{l=0}^{m-1} {(-1)^l \over l! (m-l)!} \Delta_{i_1...i_{m-l}} \sum_{j_1,
  j_2, ..., j_l} I_{i_1i_2...i_{m-l} j_1...j_{l}} ~.
\ee
In the limit $\beta_i \to 0$, we have 
\be
\Delta_{i_1...i_k} = \pi \gamma (1-\cos(\beta_{i_1} +...+\beta_{i_k}))= (\alpha_{i_1} + ... + \alpha_{i_k})^2 + {\cal O}(\alpha^4/\gamma).
\ee
  Since $I_{ijk...}$ is completely symmetric,
\be  \Delta_{i_1...i_{m-l}} I_{i_1i_2...i_{m-l} j_1...j_{l}} = \left(
  (m-l)\alpha_{i_1}^2 + (m-l)(m-l-1)\alpha_{i_1}\alpha_{i_2} \right)
I_{i_1i_2...i_{m-l} j_1...j_{l}}~.
\ee
  Therefore Eq. \ref{termx} is equal to
\be
\sum_{i_1,...,i_m=1}^n I_{i_1...i_m} \sum_{l=0}^{m-1}  {(-1)^l \over l!} \left( {\alpha_{i_1}^2 \over (m-l-1)!} + {\alpha_{i_1} \alpha_{i_2} \over (m-l-2)! } \right) = \sum_{i_1,...,i_m=1}^n I_{i_1...i_m} \left( \alpha_{I_1}^2 \delta_{m,1} + \alpha_{i_1} \alpha_{i_2} \delta_{m,2} \right).
\ee
So only the 1- and 2-disk overlap areas contribute!   Therefore the exponent in Eq. \ref{general2} becomes simply
\be
-\alpha_i^2 I_i - \alpha_i \alpha_j I_{ij}.
\ee
Recalling that $I_i = \ln(R/\delta)$ and $I_{ij} = \ln(R/|z_{ij}|)(1-\delta_{ij})$, and using $\sum_i \alpha_i^2 = \sum_{ij} \left( \alpha_i \alpha_j - \alpha_i \alpha_j (1-\delta_{ij}) \right)$, we get
\be
\la e^{i \beta_1 N(z_1)} e^{i \beta_2 N(z_2)} ... e^{i \beta_n N(z_n)} \ra =
\left| \exp \left\{ -\sum_{i,j} \alpha_i \alpha_j \left( \ln {R \over \delta} + \ln {\delta \over |z_{ij}|} (1-\delta_{ij}) \right) \right\} \right|^2.
\ee
As usual this is IR divergent unless we impose conservation of charge, which here is simply $\sum_i \alpha_i =\sum_{i,j}\alpha_i \alpha_j=0$.  Then, recalling that $\beta N(z) = \sqrt{2} \alpha \phi(z)$
\be
\la e^{i \sqrt{2} \alpha_1 \phi(z_1)} e^{i \sqrt{2} \alpha_2 \phi(z_2)} ... e^{i \sqrt{2} \alpha_n \phi(z_n)} \ra = \prod_{i<j}\left|{z_i-z_j \over \delta} \right|^{4 \alpha_i \alpha_j}.
\ee
This is the general form of a correlation function of vertex operators $e^{i \sqrt{2} \alpha \phi(z)}$ of a free massless field in $D=2$ (see e.g. \cite{cft}, p. 296).  In higher dimensions a non-interacting massless scalar with non-canonical kinetic term $\phi \Box^{d/2} \phi$ could produce such correlators.

\section{Conclusions}

Starting from eternal de Sitter space we have successfully defined a model with correlation functions that are conformally covariant and transform with positive weight.  However because of the lack of analyticity in the 4-point function, the model does not appear to be a healthy conformal field theory (except perhaps in the non-interacting $\gamma \to \infty$ limit).

One possibility is that the theory we have defined here is the limit of some good CFT in which certain effects have been ignored.  Such limits can result in non-analyticities in correlation functions (for example, one can get logs from power laws by expanding around a limit where the dimensions of some operators go to zero).  Adding weight to this possibility is that in defining the simplest possible non-trivial model, we indeed have ignored many potentially important effects:
\begin{itemize}
\item We have ignored perturbative fluctuations of the field around its minima, and included only the instantons.
\item We have ignored perturbative corrections to the instantons themselves, which for example include fluctuations away from spherical shape.
\item We have ignored interactions between the instantons other than their collisions, and we have treated collisions and overlaps in a simplistic manner.
\item We have used a semi-classical approximation that ignores quantum interference between different configurations in the ensemble of bubbles.
\item We have ignored gravitational fluctuations in the bulk, as well as fluctuations in the geometry of the boundary.
\item The bulk theory we considered is a single scalar field in de Sitter space.  Even coupled to gravity, such a model in anti-de Sitter space probably does not define a consistent CFT---one presumably needs the infinite number of modes of string theory, or at least one expects some very restrictive conditions on the bulk degrees of freedom necessary to define a good dual.  Perhaps similar restrictions apply here.
\end{itemize}

From this point of view the model described here is only a potentially interesting first step on the path to a full theory of inflation or de Sitter space.  Nevertheless, we feel the basic structure may be correct, and indeed many of the effects mentioned above could be included as perturbations around our limit.  In addition, the model may have interested applications to condensed matter systems or as an example of a new class of conformally invariant theories.  We hope to investigate some of these issues further in the future.

\section*{Acknowledgements}

We would like to thank Raphael Bousso, Paul
Goldbart, Alexander Grossberg, Andrei
Gruzinov, Joanna Karczmarek, Xiao Liu, Juan Maldacena, Yu Nakayama,
Alberto Nicolis, Massimo Porrati, Kris Sigurdson, Dan Stein, Lenny
Susskind, I-Sheng Yang,  Alex Vilenkin, and Bob Ziff for discussions.
We are especially grateful to Gaston Giribet, Simeon Hellerman, Peter Kleban, and Steve Shenker for very helpful
conversations.  The work of MK is supported by NSF CAREER grant
PHY-0645435. 
BF is supported by the Berkeley Center for 
Theoretical Physics, by a CAREER grant (award number 0349351) of the National Science Foundation, and by 
the US Department of Energy under Contract DE-AC02- 
05CH11231. 
 MK and BF would like to thank the Aspen Center for
Physics and the Banff International Research Station, where this work
was initiated, for their hospitality.

\bibliographystyle{utphys}

%\bibliography{when}

\end{document}